\documentclass[fleqn,usenatbib]{mnras}

\usepackage{newtxtext,newtxmath}
\usepackage[T1]{fontenc}
\usepackage{lineno}
\DeclareRobustCommand{\VAN}[3]{#2}
\let\VANthebibliography\thebibliography
\def\thebibliography{\DeclareRobustCommand{\VAN}[3]{##3}\VANthebibliography}

\usepackage{graphicx}	% Including figure files
\usepackage{amsmath}	% Advanced maths commands
\usepackage{color}
\usepackage{pdflscape}

\newcommand\hxmt{{\it Insight}-HXMT}
\title[A mHz QRM study of 4U 1630-47]{The mHz quasi-regular modulations of 4U 1630--47 during its 1998 outburst}
\author[Q. C. Zhao et al.]{Q. C. Zhao$^{1, 2, 3}$, 
H. X. Yin$^{1}$\thanks{E-mail: yinhx@sdu.edu.cn}, 
L. Tao$^{2}$, 
Z. X. Yang$^{2, 3}$, 
J. L. Qu$^{2}$, 
L. Zhang$^{2}$, 
S. Zhang$^{2}$, 
E. L. Qiao$^{5, 6}$,
\newauthor{ Q. C. Bu$^{7}$, S. J. Zhao$^{2, 3}$, P. P. Li$^{2, 3}$, Y. M. Huang$^{2, 3}$, R. C. Ma$^{2, 3}$, R. J. Tang$^{4}$, P. Jin$^{2, 3}$, W. Yu$^{2, 3}$ }
\newauthor{H. X. Liu$^{2, 3}$, Y. Huang$^{2}$, X. Ma$^{2}$, J. Y. Xiao$^{2, 3}$, X. Zhang$^{2}$, K. Zhao$^{2, 3}$}
\\
$^{1}$Shandong Key Laboratory of Optical Astronomy and Solar-Terrestrial Environment, School of Space Science and Physics, Institute of Space Sciences,\\
Shandong University, Weihai, Shandong, 264209, China\\
$^{2}$Key Laboratory of Particle Astrophysics, Institute of High Energy Physics, Chinese Academy of Sciences, 100049 Beijing, China\\
$^{3}$Uinversity of Chinese Academy of Sciences, Chinese Academy of Sciences, 100049 Beijing, China\\
$^{4}$ Beijing Jiaotong University, Beijing 100044, China \\
$^{5}$ Key Laboratory of Space Astronomy and Technology, National Astronomical Observatories, Chinese Academy of Sciences, Beijing 100101, China \\
$^{6}$ School of Astronomy and Space Sciences, University of Chinese Academy of Sciences, 19A Yuquan Road, Beijing 100049, China \\
$^{7}$Institut f\"ur Astronomie und Astrophysik, Kepler Center for Astro and Particle Physics, Eberhard Karls Universit\"at, Sand 1, D-72076 T\"ubingen, Germany\\}
\date{Accepted XXX. Received YYY; in original form ZZZ}
\pubyear{2015}
\begin{document}
\label{firstpage}
\pagerange{\pageref{firstpage}--\pageref{lastpage}}
\maketitle
% Abstract of the paper

\begin{abstract}
We present the results of a detailed timing and spectral analysis of the quasi-regular modulation (QRM) phenomenon in the black hole X-ray binary 4U 1630--47 during its 1998 outburst observed by Rossi X-ray Timing Explore (RXTE).  We find that the $\sim$ 50-110 mHz QRM is flux dependent, and the QRM is detected with simultaneous low frequency quasi-periodic oscillations (LFQPOs). According to the behavior of the power density spectrum, we divide the observations into four groups. In the first group, namely behavior A, LFQPOs are detected, but no mHz QRM. The second group, namely behavior B, a QRM with frequency above $\sim$ 88 mHz is detected and the $\sim$ 5 Hz and $\sim$ 7 Hz LFQPOs are almost overlapping. In the third group, namely behavior C, the QRM frequency below $\sim$ 88 mHz is detected and the LFQPOs are significantly separated. In the forth group, namely behavior D, neither QRM nor LFQPOs are detected. We study the energy-dependence of the fractional rms, centroid frequency, and phase-lag of QRM and LFQPOs for behavior B and C. We then study the evolution of QRM and find that the frequency of QRM increases with hardness, while its rms decreases with hardness. We also analyze the spectra of each observation, and find that the QRM rms of behavior B has a positive correlation with  $\rm F_{\rm powerlaw}$ / $\rm F_{\rm total}$. Finally, we give our understanding for this mHz QRM phenomena.
\end{abstract}

% Select between one and six entries from the list of approved keywords.
% Don't make up new ones.
\begin{keywords}
black hole physics – accretion, accretion disc – binaries (including multiple): close – X-rays: binaries – X-rays: individual (4U 1630-47)
\end{keywords}

%%%%%%%%%%%%%%%%%%%%%%%%%%%%%%%%%%%%%%%%%%%%%%%%%%

%%%%%%%%%%%%%%%%% BODY OF PAPER %%%%%%%%%%%%%%%%%%

\section{Introduction}

A black hole X-ray binary system consists of a black hole and a companion star. The black hole of low mass X-ray binary can accrete matter from the companion star through the Roche lobe and enter an outburst state. During this process, the luminosity will increase by several orders of magnitude. For a canonical outburst, it follows a "q"-shape evolution process on hardness-intensity diagram (HID) \citep{homan_evolution_2005, fender_jets_2009, 2016ASSL..440...61B}, going through: low hard state (LHS), hard intermediate state (HIMS), soft intermediate state (SIMS), high soft state (HSS), and then returns back to SIMS, HIMS, LHS, and finally enters quiescence. During outbursts, black hole X-ray binaries will show a significant evolution both with respect to their energy spectra and their timing properties \citep{1994ApJS...92..511V, 1997ApJ...479..926M}. The X-ray spectra of Black hole candidates (BHC) consists of two components: a thermal and a non-thermal component. The thermal component originates from accretion disc \citep{shakura_black_1973}, and the seed photons from the disc are scattered by the corona or the base of jet, forming the non-thermal component \citep{yuan_hot_2014}. Part of the photons emitted by the corona or the base of the jet will illuminate on the disc, forming a reflection component \citep{2014ApJ...782...76G, 2015ApJ...808L..37G}.

In the time domain, studying the X-ray variability behavior on short timescales can help us better understand the process and geometry of accretion flow \citep{belloni_fast_2014}. The Fourier transform is a very useful tool when studying the behavior of X-ray variability \citep{van_der_klis_fourier_1989,ingram_review_2019}. Low frequency quasi-periodic oscillation (LFQPOs) are among one of the most remarkable timing features on the power density spectrum (PDS) of black hole X-ray binaries  \citep{nowak_are_2000, belloni_unified_2002, van_der_klis_qpo_2005, ingram_review_2019}. LFQPOs can be divided into three categories according to their centroid frequency, time/phase-lag, and total RMS amplitude: type-A, B, and C \citep{homan_evolution_2005, casella_abc_2005, ingram_review_2019}. The evolution of LFQPOs is related to the energy spectral states. Type-C QPOs are mainly observed during the LHS and HIMS. The appearance of type-B QPOs mark the entry into the SIMS, while the type-A QPOs mainly appears in the HSS. LFQPOs are of great significance for us to understand the structure of accretion flows and the accretion processes in extreme environments. Although many models have been proposed to explain LFQPOs \citep{ingram_low-frequency_2009, ingram_physical_2011, ma_discovery_2021}, the origin of LFQPOs is very much still under debate.

The study of the energy dependence of the LFQPO frequency and the RMS amplitude is also important for us to comprehend the accretion processes around black holes. The LFQPO RMS increases with energy below about 10 keV, then flattens out above 10 keV, and may increase or decrease at higher energies \citep{qu_energy_2010, zhang_z_variability_2014, 2020ApJ...895..120L}. \citet{2018ApJ...866..122H} found the fractional rms increases with energy up to  $\sim$ 20 keV, then flattens in the range of 20--100 keV with Insight-HXMT observations of MAXI J1535-571. In addition to LFQPOs, quasi-regular modulations have also been observed in several sources, such as: 4U 1630--47 \citep{trudolyubov_rxte_2001, 2022ApJ...937...33Y},  H 1743--322 \citep{altamirano_low-frequency_2012},  GRS 1915+105 \citep{morgan_lessigreaterrxtelessigreaterobservations_1997}, and GRO J1655--40 \citep{remillard_lessigreaterrxtelessigreaterobservations_1999}. The frequency of these modulations in the PDS is in the range of several to tens mHz, such as the "heartbeat" QPOs observed in GRS 1915+105 \citep{2000A&A...355..271B,2002MNRAS.331..745K}. These "heartbeat" QPOs may have an association with unstable disc limit cycles of accretion and ejection \citep{2011ApJ...737...69N}. The frequency of "heartbeat"  QPOs is in the mHz range, and appears in the high luminosity state.

4U 1630--47 is a very interesting black hole source, which was first observed by the Vela-5B satellite in 1969 \citep{priedhorsky_recurrent_1986}. It has exhibited frequent outbursts \citep{choudhury_qpo_2015} since its discovery, with more than 20 outbursts observed so far \citep{tetarenko_watchdog_2016}, undergoing an outburst about every 600-700 days \citep{priedhorsky_recurrent_1986, parmar_periodic_1995}. Since the optical or infrared counterpart of the source has not been identified, no dynamical measurements of the mass can be made. However, \citet{seifina_black_2014} estimated that the mass of the source is about 10 $~\rm M_\odot$ and the inclination is $\leq$ 70 deg by the scaling of the correlation of the photon index with low frequency quasi-periodic oscillations. The presence of a dip observed in the 1996 outburst indicates that the source has a high inclination \citep{tomsick_x-ray_1998}. \citet{king_disk_2014} fitted the energy spectra with {\it NuSTAR} using the reflection model {\sc refbhb}, and obtained an inclination of $\thicksim$ 64 deg and spin parameter $a_\ast$  of $\thicksim$ 0.985. \citet{2022MNRAS.512.2082L} obtained a moderate spin, $\thicksim$ 0.817, but a very low inclination $\thicksim$ 4 deg by fitting the \hxmt\ data. The distance is estimated to be around 10 $\sim$ 11 kpc \citep{seifina_black_2014}.

For 4U 1630--47, most of the outbursts are not observed in the hard state; this may be due to the fast transition time from the hard state to soft state which our instruments cannot capture in time \citep{capitanio_missing_2015, baby_astrosat_2020}, a similar problem that is exhibited by MAXI J0637--430 \citep{2022MNRAS.514.5238M}. Since the 1998 outburst was relatively well documented, many of the timing features of this outburst have been studied. \citet{dieters_timing_2000} studied the timing evolution of the 1998 outburst, and found that the light curve of a special behavior was gradually dominated by sharp dips. \citet{tomsick_x-ray_2000} presented a study of spectral and timing analysis of the 1998 outburst and found that a soft-to-hard state transition occurred 104 days after the start of the outburst. \citet{trudolyubov_rxte_2001} divided the outburst into three plateaus, with the flux of each individual plateau roughly constant and the timing and energy spectra properties relatively stable and significantly different from each of the other plateaus. The mhz QRM mainly appeared in plateau B. 

\citet{trudolyubov_rxte_2001} reported a QRM phenomenon during the 1998 outburst. Recently, \citet{2022ApJ...937...33Y} reported a mHz QRM phenomenon during the 2021 outburst of 4U 1630--47 with \hxmt. They found that the fractional rms of mHz QRM has a positive correlation with photon energy in the range of 1--100 keV. They explain this phenomenon with an instability model of accretion rate dependence. In contrast to the 2021 outburst, the QRM during the 1998 outburst is detected with simultaneous LFQPOs. While the QRM phenomenon has been reported, the properties of mHz QRM and LFQPOs during the 1998 outburst of 4U 1630--47 have not been thoroughly explored through detailed timing analysis via (1) dynamical power density spectrum; (2) frequency-resolved phase-lag spectrum; (3) energy-dependent properties of the centriod frequency, fractional rms, and phase-lag. It would be interesting to compare the mHz QRM phenomenon seen in the 1998 outburst and the 2021 outburst. In this paper, we present a detailed timing and spectral analysis of the mHz QRM phenomenon exhibited by 4U 1630-47 during the 1998 outburst. In section 2, we describe the data used in our data analysis and the processing methods. We present the main results in Section 3 and the discussion in Section 4.

\section{OBSERVATIONS AND DATA REDUCTION}
 We process RXTE/PCA data with HEAsoft version 6.24. The criteria of generating the good time intervals (GTIs) is an elevation angle larger than $10^{\circ }$ and offset less than $0.02^{\circ}$. The standard data products, e.g. light curves, spectra, and background are produced from the Proportional Counter Unit 2 (PCU2) of RXTE/PCA. We use XSPEC version 12.10.0 to perform the spectral analysis \citep{1996ASPC..101...17A}. The energy band for our spectral analysis is 2.5--30.0 keV. We add 0.5 $\%$ systematic error, in order to account for the calibration uncertainties \citep{2021MNRAS.508..287S}.  Uncertainties (90$\%$-confident-level) are calculated by Markov Chain Monte Carlo and the chain length is 20000. Binned mode and Event mode data are used for our timing analysis.
 
\begin{figure}
    \centering
    \includegraphics[width=0.4\textwidth]{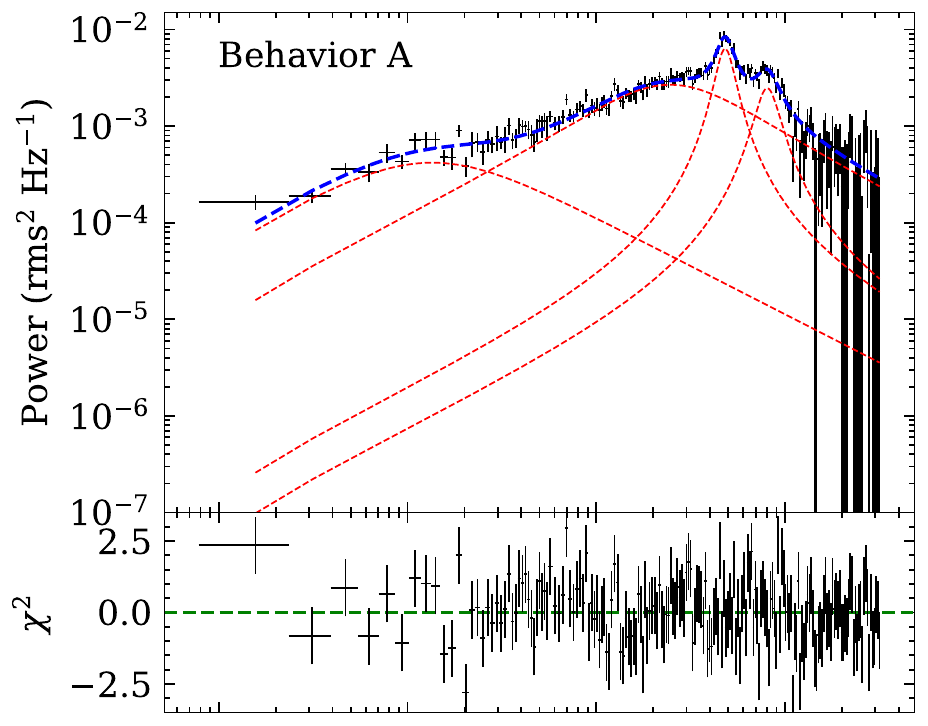}
    \includegraphics[width=0.4\textwidth]{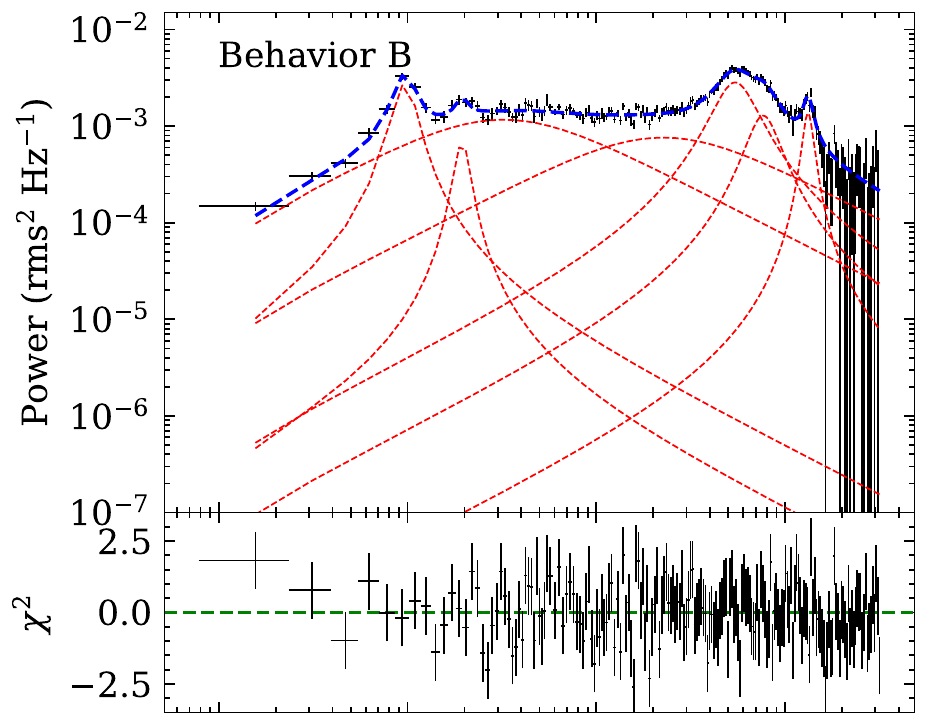}
    \includegraphics[width=0.4\textwidth]{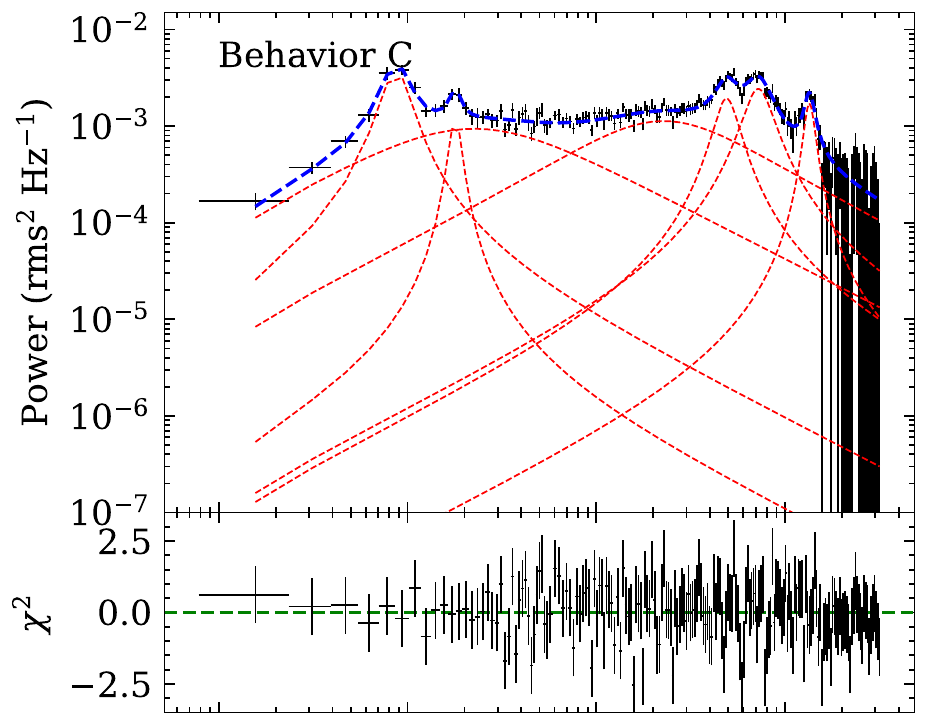}
    \includegraphics[width=0.4\textwidth]{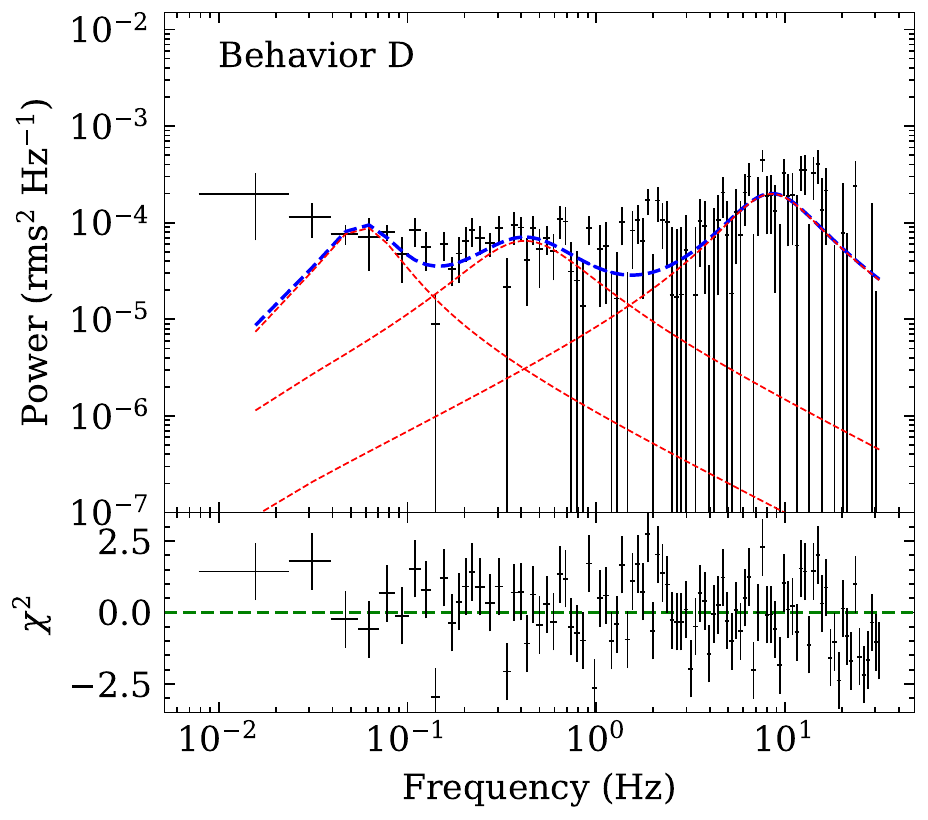}
    \caption{Four representative PDS for behavior A, B, C and D. The corresponding observations are 30188-02-04-00, 30178-02-02-01, 30188-02-11-00 and 30188-02-12-00, respectively.}
    \label{Fig:Fig1}
\end{figure}

 We generate the power density spectra (PDS) with 1/64 s resolution and 64 s long interval to find observations with mHz QRMs. The power density spectum is normalized with miyamoto normalization \citep{1991ApJ...383..784M}. We fit the PDS with multiple Lorentzian functions \citep{belloni_unified_2002}. The obtained fractional rms is background-corrected by the formula: $~\rm rms= ~\rm \sqrt{P}\times(S+B)/S$ \citep{belloni_atlas_1990, 2015ApJ...799....2B, kong_joint_2020}.  As shown in Fig. \ref{Fig:Fig1}, according the features in PDS, we define four different behavior PDS: behavior A: no QRM but LFQPOs are detected; behavior B: the QRM frequency above $\sim$ 88 mHz is detected and the $\sim$ 5 Hz and $\sim$ 7 Hz LFQPOs are almost overlapping; behavior C: the QRM frequency below $\sim$ 88 mHz is detected and the $\sim$ 5 Hz and $\sim$ 7 Hz LFQPOs are significantly separated; behavior D: neither QRMs nor LFQPOs are detected.

\begin{figure}
    \centering
    \includegraphics[width=0.45\textwidth]{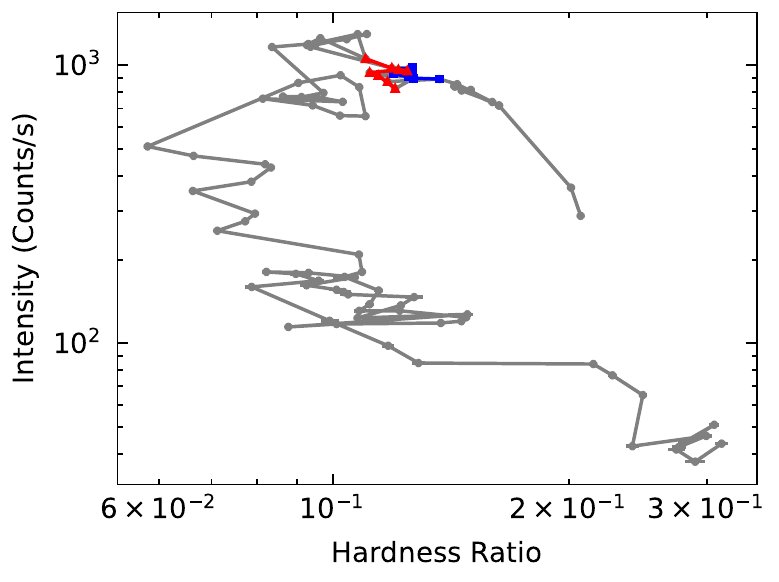}
    \includegraphics[width=0.45\textwidth]{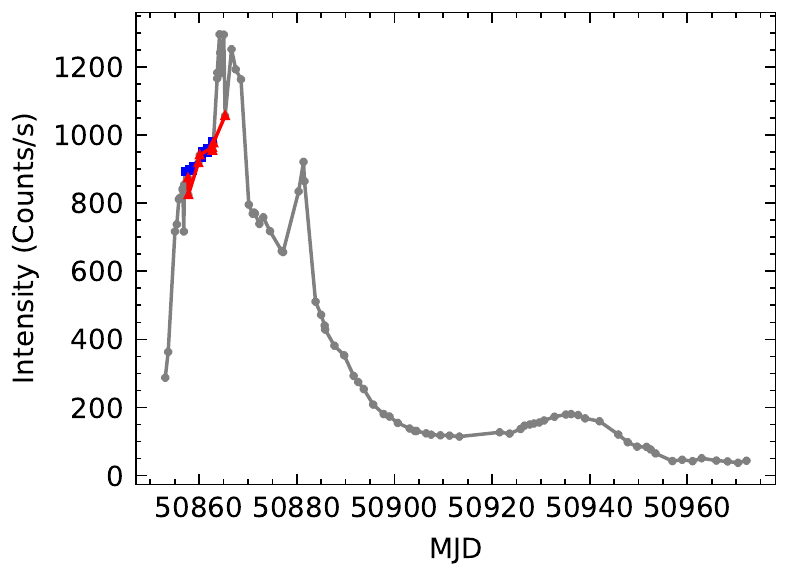}
    \caption{Top panel: the hardness-intensity diagram of 4U 1630--47 in the 1998 outburst with RXTE/PCA (PCU2). Hardness is defined as the count rates ratio between 1.94--11.91 keV and 11.91--30.02 keV. Bottom panel: the evolution of lightcurve (1.94--30.02 keV) over time during the 1998 outburst. Blue squares represent the observations of behavior B.  Red triangles represent the observations of behavior C. For the definition of behavior B and C, see below for details.}
    \label{Fig:Fig2}
\end{figure}

\section{Results}
\subsection{Fundamental diagrams}

In Fig. \ref{Fig:Fig2}, we show the HID and light curve of the 1998 outburst, using RXTE/PCA data. The hardness is defined as the count rates ratio between 1.94--11.91 keV and 11.91--30.02 keV over time during 1998 outburst. This outburst shows a transition from the hard state to the soft state. The mHz QRM was first observed at MJD 50857.12 (observation is 30178-01-05-00) and last observed at MJD 50865.31 (observation is 30188-02-16-00). Based on the position at the HID, we can roughly judge that the mHz QRM appears in the intermediate state.

\begin{figure*}
    \centering
    \includegraphics[width=0.8\textwidth]{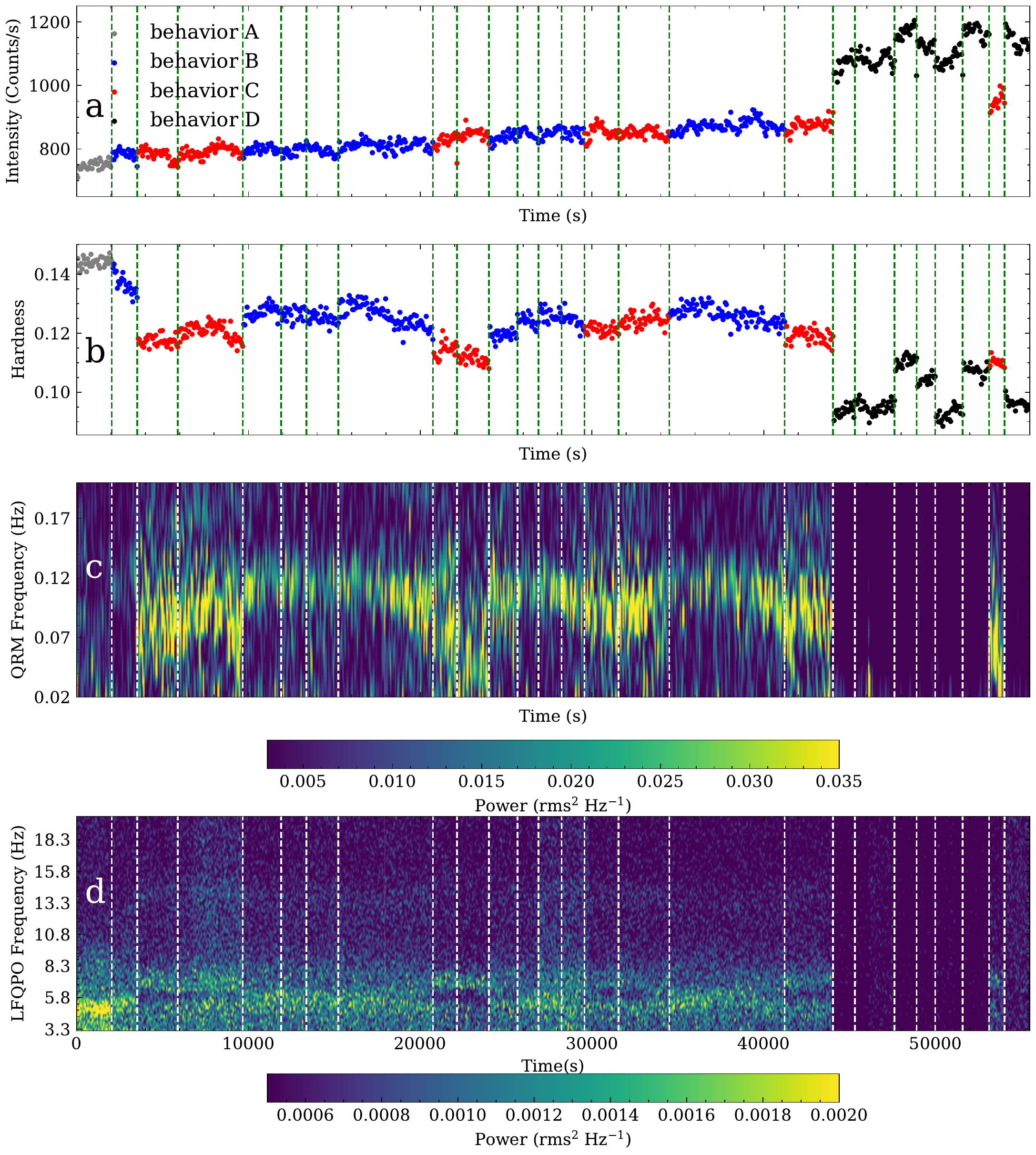}
    \caption{Panel a: the lightcurve of PCU2 (2--13 keV). Panel b: the hardness evolution. Hardness is defined as the count rates ratio between 1.94--11.91 keV and 11.91--30.02 keV (PCU2). Panel c: the dynamic PDS of mHz QRM. Panel d: the dynamic PDS of LFQPOs. The dash lines represents the gap between different observations. The brighter color indicates the greater power on the dynamical PDS. The grey dots, blue dots, red dots and black dots represent behavior A, B, C, and D, respectively.}
    \label{Fig:Fig3}
\end{figure*}

In Fig. \ref{Fig:Fig3}, we show the dynamic power spectra of mHz QRM and LFQPOs (panel c and panel d). From panel a and panel b in Fig. \ref{Fig:Fig3}, we can see that the mHz QRM seems to only appear in a specific flux range, and when the flux suddenly increases, the mHz QRM disappears. In particular, when the flux returns back to the specified range, the mHz QRM reappears. The hardness ratio remains relatively stable during the mHz QRM time zones. From panels c and d in Fig. \ref{Fig:Fig3}, we can see that the $\sim$ 7 Hz LFQPO is more apparent when the frequency of mHz QRM is lower and the width of mHz QRM is wider. When the $\sim$ 7 Hz LFQPO is obvious, the hardness seems lower than that of other segments with mHz QRM detected. 
\subsection{Timing analysis}

\begin{figure}
    \centering
    \includegraphics[width=0.40\textwidth]{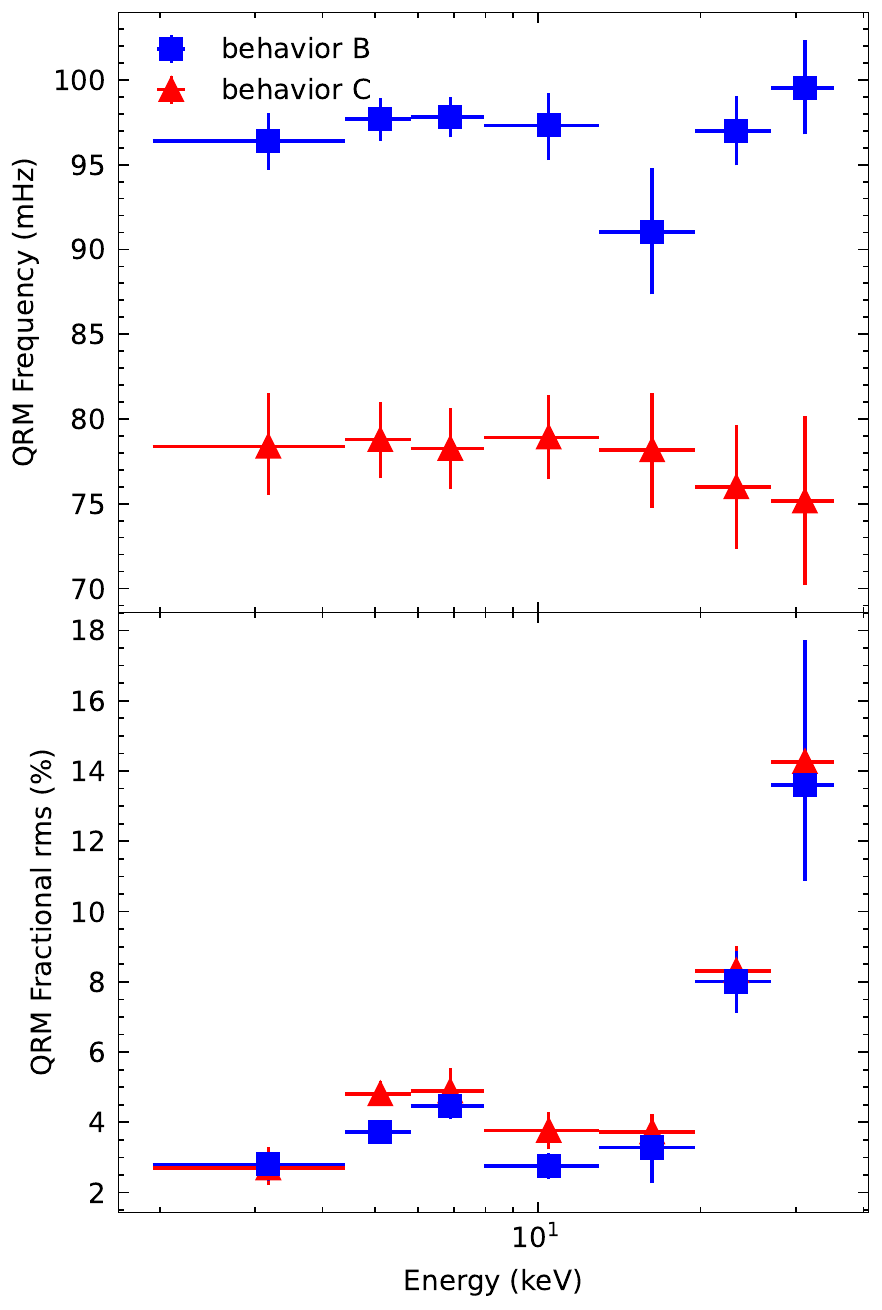}
    \caption{Top panel: the centroid frequency of the mHz QRM as a function of photon energy. Bottom panel: the fractional rms of the mHz QRM as a function of photon energy. The blue squares corresponds to behavior B PDS, and the red triangles corresponds to behavior C PDS. }
    \label{Fig:Fig4}
\end{figure}

\subsubsection{Energy-dependence}
\begin{figure}
    \centering
    \includegraphics[width=0.4\textwidth]{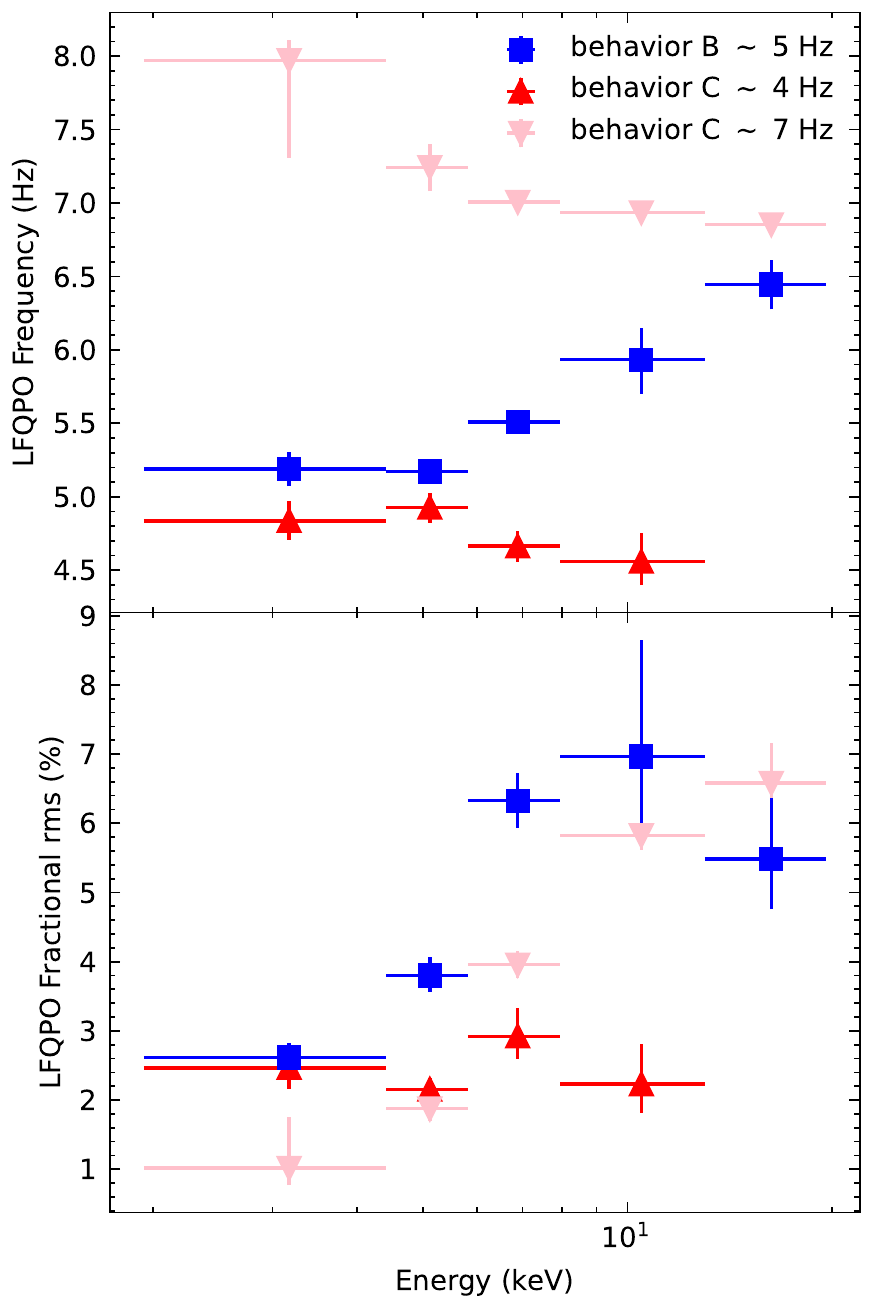}
    \caption{Top panel: the centroid frequency of the LFQPO as a function of photon energy. Bottom panel: the fractional rms of the LFQPO as a function of photon energy. The blue squares, the read triangles, and the cyan inverted triangles corresponds to the LFQPO ($\sim $ 5 Hz) of behavior B,  LFQPO ($\sim$ 4 Hz) of behavior C, and  LFQPO ($\sim$ 7Hz) of behavior C, respectively}
    \label{Fig:Fig5}
\end{figure}

We combine observations based on the behavior divisions in Table \ref{table: table1} to improve the signal/noise ratio. We divide multiple energy bands (1.94--4.41 keV, 4.41--5.82 keV, 5.82--7.96 keV, 7.96--12.99 keV, 12.99--19.56 keV, 19.56--27.00 keV, 27.00--35.35 keV)  to study the energy dependence of the centroid frequency and the fractional rms of the mHz QRM and LFQPOs. In Fig. \ref{Fig:Fig4} top panel, we show the frequency of QRM as a function of energy. The QRM frequency for behavior B is higher than that for behavior C, but the evolution trend with photon energy is similar: roughly equal in different energy bands. The unusual dot of 12.99 keV--19.56 keV for behavior B may be caused by a slightly lower signal/noise ratio.  From the bottom panel of Fig. \ref{Fig:Fig4}, we can see that the QRM fractional rms of behavior B and C show a similar trend: below 13 keV, there is a slight hump between 5--7 keV; above 13 keV, QRM fractional rms increases with energy. For the 2021 outburst, \citet{2022ApJ...937...33Y} found the fractional rms of mHz QRM has a positive correlation with photon energy. Due to the limitation of the PCA energy band, we cannot give the energy-dependent properties at higher energy bands, but considering the energy bands covered by the two outbursts, they show relatively similar properties. In the top panel of Fig. \ref{Fig:Fig5}, we show the LFQPOs centroid frequency as a function of photon energy of behavior B and behavior C. The LFQPO ($\sim$ 5 Hz) frequency of behavior B increases with energy. But LFQPO ($\sim$ 4 Hz) of behavior C seems to be equal in different energy bands. The LFQPO ($\sim$ 7 Hz) frequency of behavior C decreases with  energy. In the bottom panel of Fig. \ref{Fig:Fig5}, LFQPO ($\sim$ 5 Hz) fractional rms of behavior B increases with energy below $\sim$ 10 keV, and slightly decreases with energy above $\sim$ 10 keV. The LFQPO fractional rms of behavior C ($\sim$ 4 Hz) seems to be equal at different energy bands and the LFQPO ($\sim$ 7 Hz) fractional rms of behavior C increases with energy. In last two or three energy bins, LFQPOs are not significantly detected.

\subsubsection{Phase-lag spectra}
Phase-lags are of great significance for studying the behavior of X-rays variability and the accretion geometry of black holes. We  calculate the phase-lag of the mHz QRM with a reference energy range of 1.94--4.41 keV. To quantitatively analyze the phase-lag behavior, we calculate the average lag in a range of $\nu\pm \rm FWHM/2$, where $\nu$ is the centroid frequency of QRM/LFQPO, and the FWHM is its full-width at half-maximum \citep{reig_phase_2000, qu_energy_2010}.

\begin{figure*}
    \centering
    \includegraphics[width=0.85\textwidth]{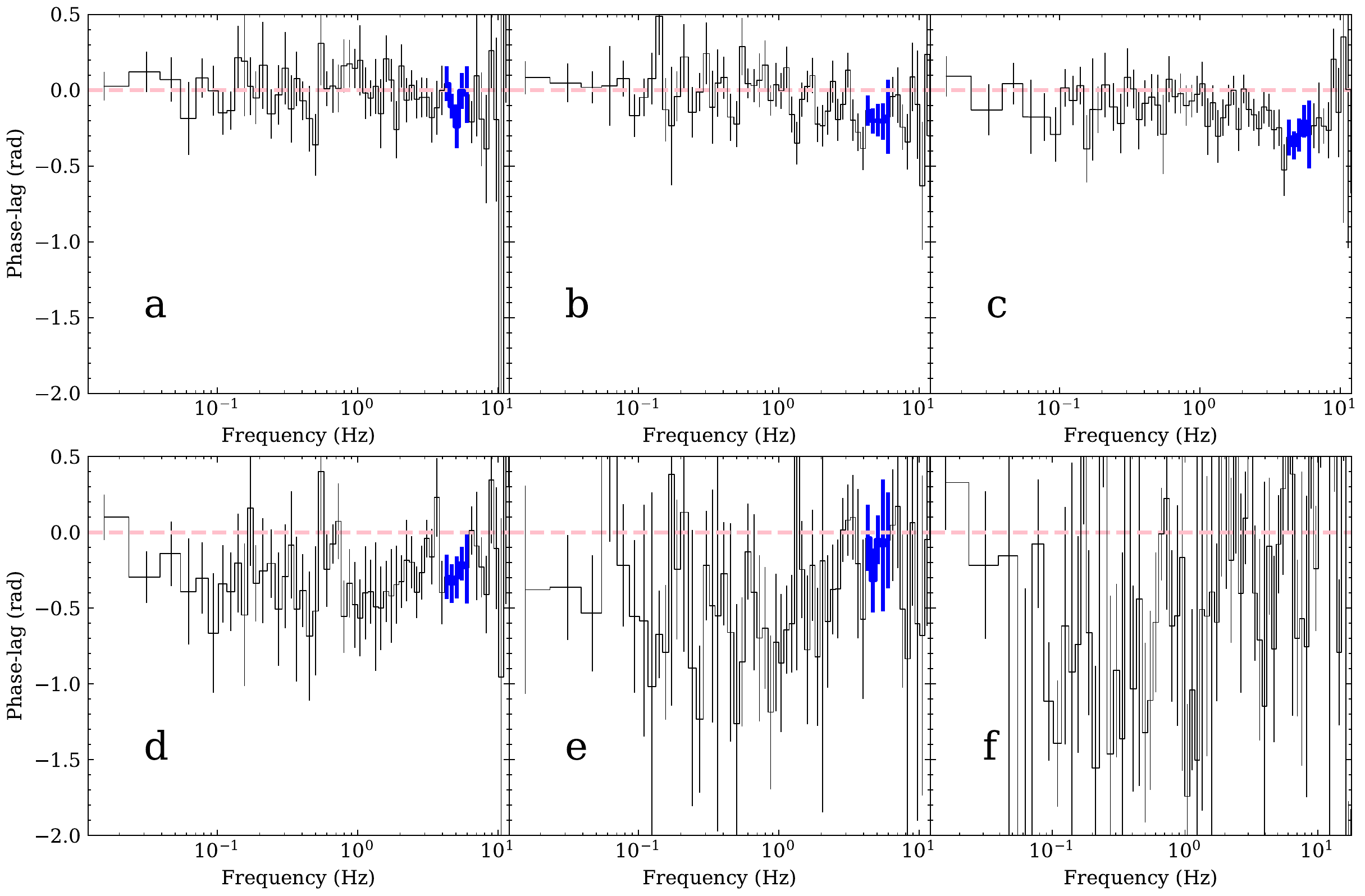}
    \caption{ Frequency-resolved phase-lag spectra in different bands of behavior A. The reference energy band is 1.94--4.41 keV. The corresponding energy band of a, b, c, d, e, and f are 4.41--5.82 keV, 5.82--7.96 keV, 7.96--12.99 keV, 12.99--19.56 keV, 19.56--27.00 keV, 27.00--35.35 keV, respectively. The blue dots represent the frequency range $\nu\pm \rm FWHM/2$ over which the phase-lags of the LFQPO are averaged. }
    \label{Fig:Fig6}
\end{figure*}

\begin{figure*}
    \centering
    \includegraphics[width=0.85\textwidth]{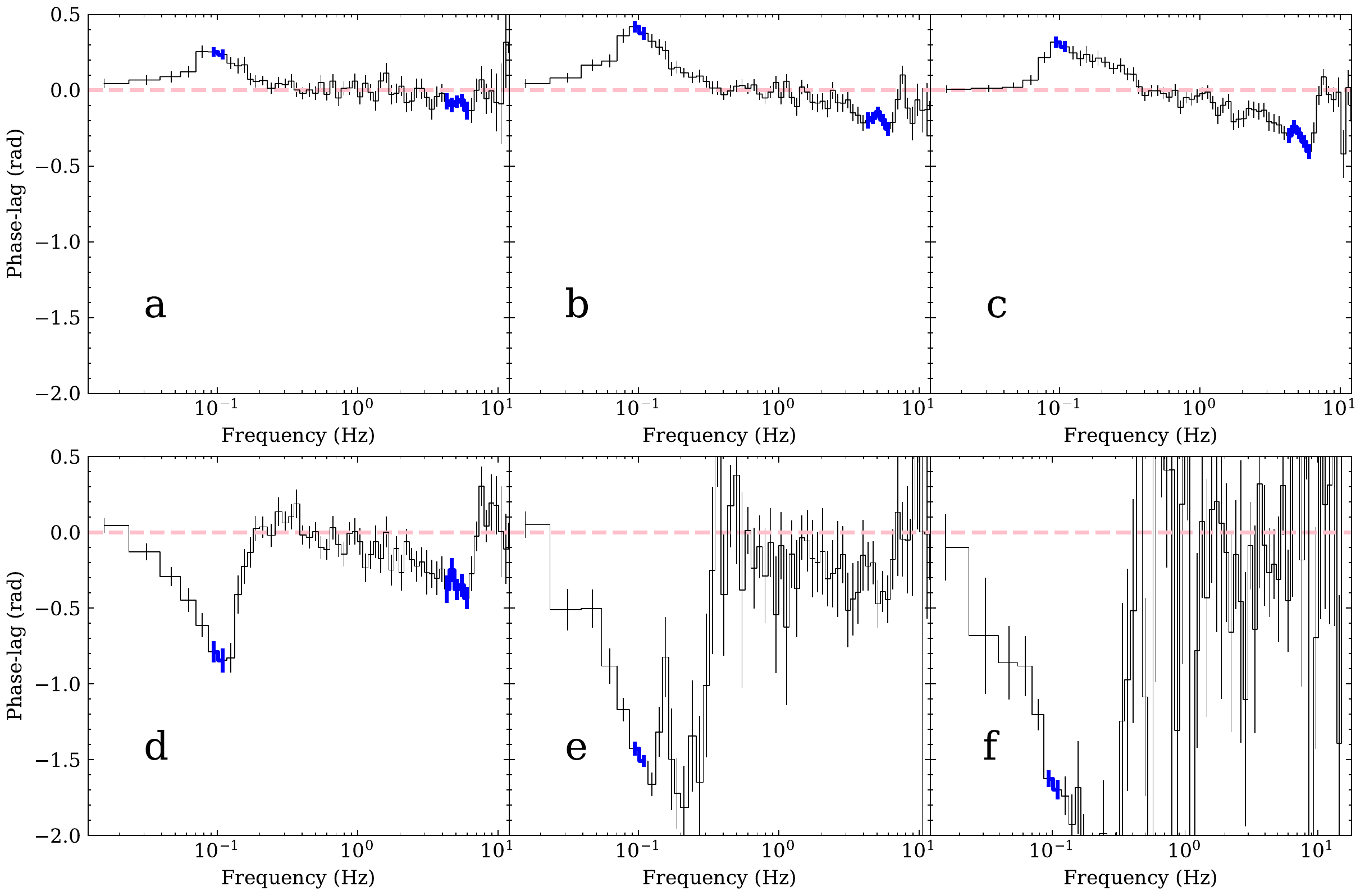}
    \caption{ Frequency-resolved phase-lag spectra in different bands of behavior B. The blue dots represent the frequency range $\nu\pm \rm FWHM/2$ over which the phase-lags of the mHz QRM/ LFQPO are averaged. }
    \label{Fig:Fig7}
\end{figure*}

\begin{figure*}
    \centering
    \includegraphics[width=0.85\textwidth]{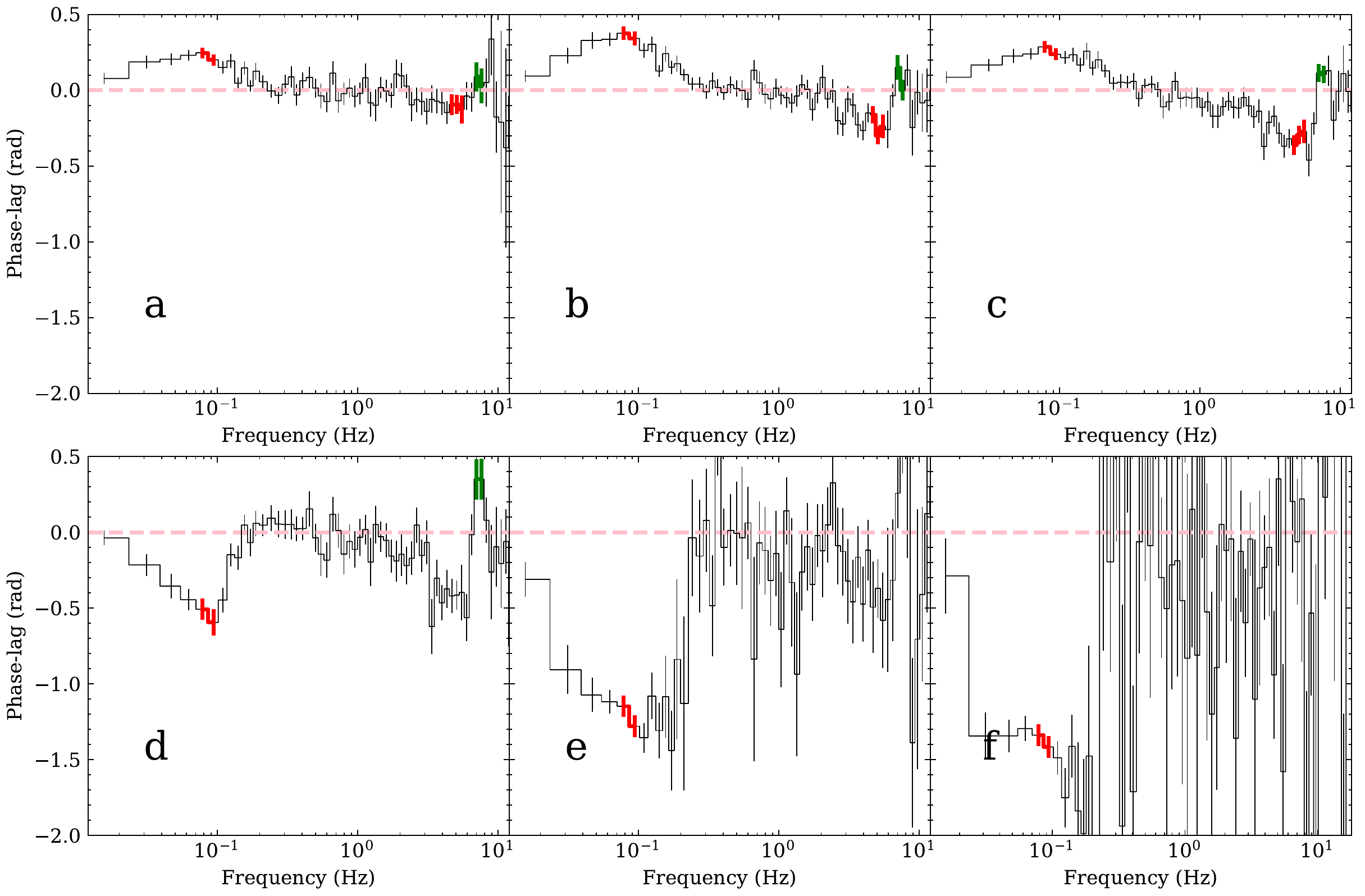}
    \caption{ Frequency-resolved phase-lag spectra in different bands of behavior C. The red and green dots represent the frequency range $\nu\pm \rm FWHM/2$ over which the phase-lags of the mHz QRM/ LFQPO are averaged.}
    \label{Fig:Fig8}
\end{figure*}

\begin{figure}
    \centering
    \includegraphics[width=0.4\textwidth]{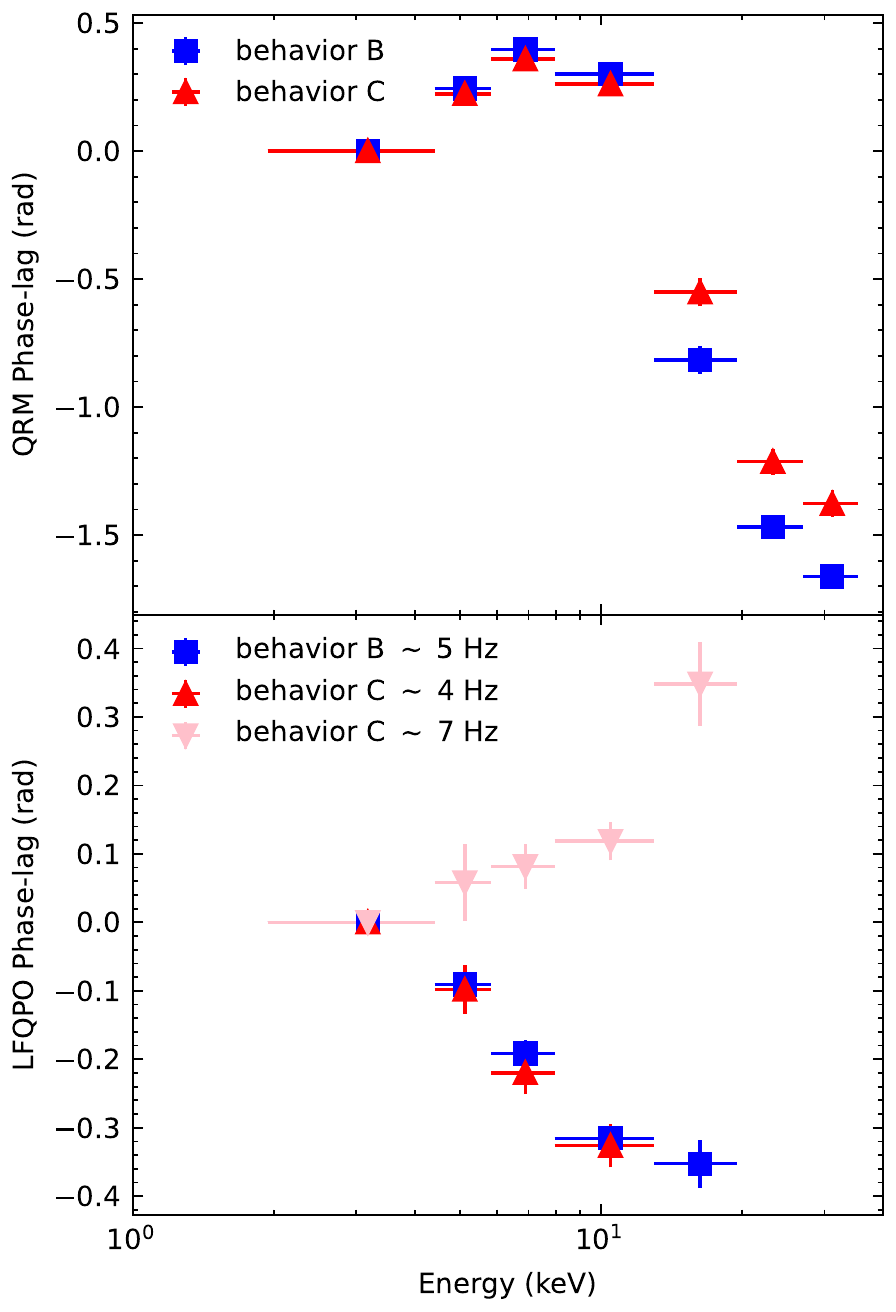}
    \caption {Top panel: phase-lag of the mHz QRM as a function of photon energy. The red triangles correspond to behavior C, the blue points correspond to behavior C. Bottom panel: Phase-lag of LFQPOs as a function of photon energy. The reference energy band is 1.94--4.41 keV. The blue squares correspond to the LFQPO ($\sim$ 5 Hz) of behavior B, the red triangles correspond to the LFQPO ($\sim$ 4 Hz) of behavior C,  and the cyan triangles represent LFQPO ($\sim$ 7 Hz)  of behavior C. The chosen bands are 4.41--5.82 keV, 5.82--7.96 keV, 7.96--12.99 keV, 12.99--19.56 keV, 19.56--27.00 keV, 27.00--35.35 keV. }
    \label{Fig:Fig9}
\end{figure}

In Fig.~\ref{Fig:Fig6}, \ref{Fig:Fig7}, and \ref{Fig:Fig8}, we show the frequency-dependent phase lag spectra of behavior A, B, and C in multiple energy bands, respectively. From the top panel of Fig.~\ref{Fig:Fig9}, we can see that mHz QRM phase-lag of behavior B and behavior C show a similar trend with energy. Below $\sim$ 13 keV, it is a slight hard lag, but when the energy exceeds $\sim$ 13 keV, it turns into a soft lag, and the soft lag increases with energy. From the bottom panel of Fig.~\ref{Fig:Fig9}, the LFQPO ($\sim$ 5 Hz) of behavior B shows a soft lag and increases with energy. The LFQPO ($\sim$ 4 Hz) phase-lag  of behavior C also shows a similar trend. But the LFQPO ($\sim$ 7 Hz) phase-lag of behavior C show a hard lag and increases with photon energy.  

\begin{figure*}
    \centering
    \includegraphics[width=0.8\textwidth]{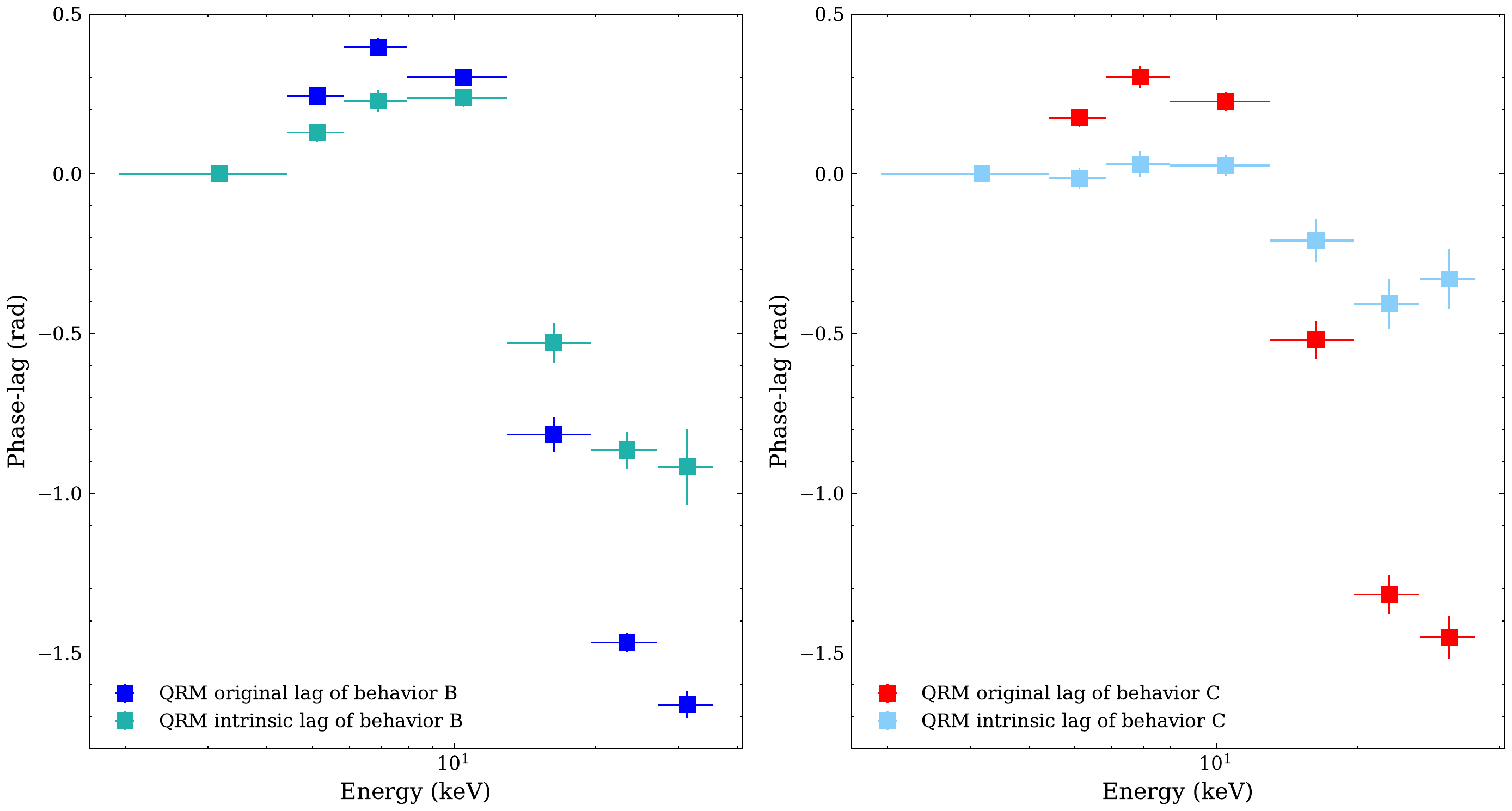}
    \caption {Left panel: phase-lag of the mHz QRM of behavior B versus photon energy, where the blue squares represent the origin lag and the lightseagreen squares represent the intrinsic lag. Right panel: phase-lag of the mHz QRM of behavior C versus photon energy, where the red squares represent the origin lag and the lightskyblue squares represent the intrinsic lag. }
    \label{Fig:Fig10}
\end{figure*}

\citet{ma_discovery_2021} describes a method for calculating the intrinsic lag when the broadband noise continuum has a significant effect on the original lag. The measurement of phase-lags of broad components in power spectra such as mHz QRMs observed in 4U 1630--47, may be influenced by the contamination from components with comparable amplitude. In order to access the impact of continuum noise component on the measurement of the phase-lags of the mHz QRMs, we calculate the intrinsic-lag following the method described by \citet{ma_discovery_2021}. It is important to note that the estimated lag is heavily influenced by the assumptions made. As depicted in Fig.~\ref{Fig:Fig10}, the intrinsic-lag ($\sim$ 0.4-0.5 rad for maximum lag of the mHz QRMs of behavior C) can be significantly smaller than the origin-lag ($\sim$ 1.5 rad for maximum lag of the mHz QRMs of behavior C), especially for the last three energy bins. Therefore, for the last three energy bins, the phase-lag of the mHz QRMs may be significantly affected by the continuum noise component.

\begin{figure*}
    \centering
    \includegraphics[width=0.8\textwidth]{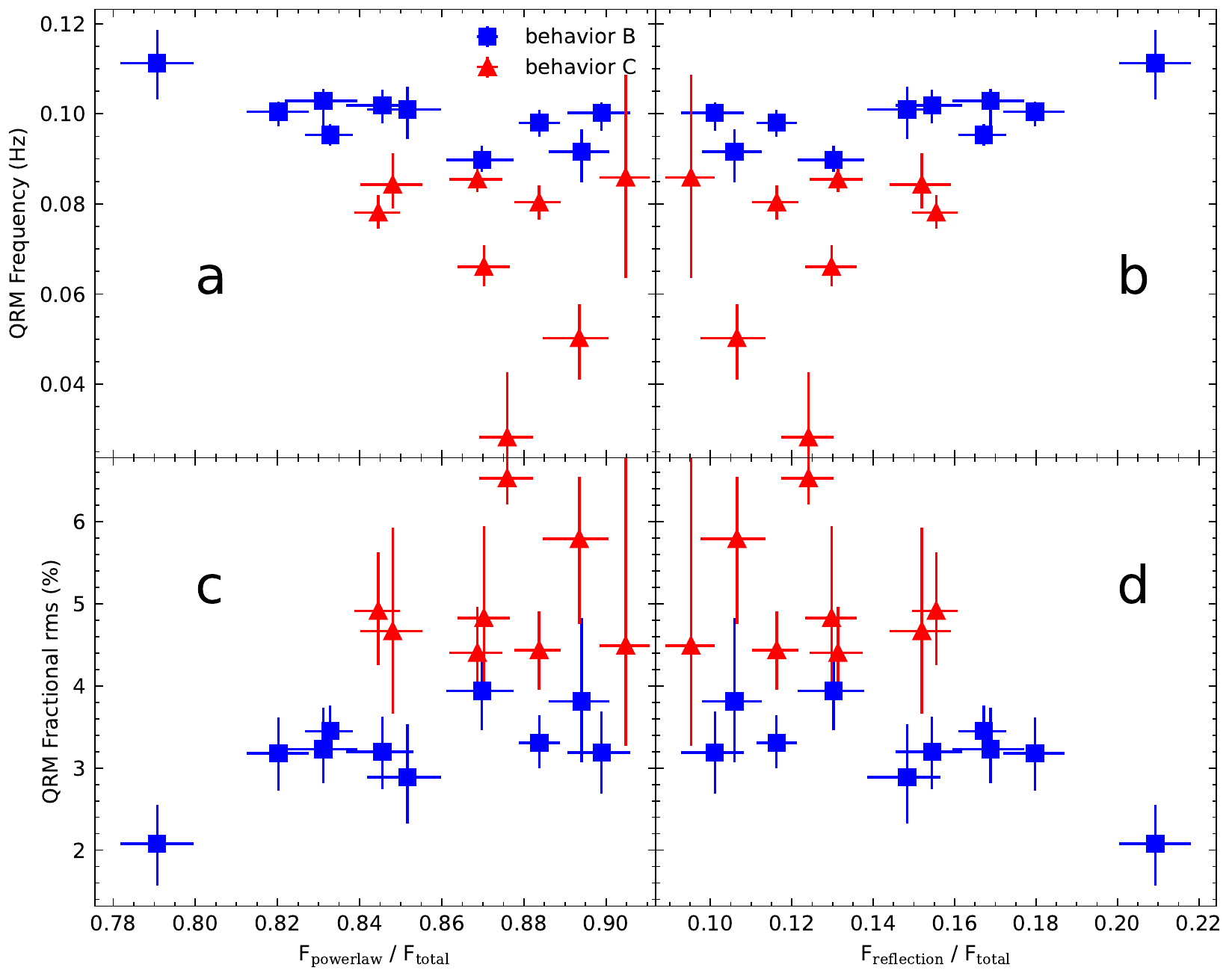}
    \caption{The mHz QRM parameters versus spectral component ﬂuxes ratio. $\rm F_{\rm toal}$ = $\rm F_{\rm powerlaw}$+$\rm F_{\rm reflection}$ . Blue squares represent the behavior B PDS. Red triangles represent the behavior C PDS.}
    \label{Fig:Fig11}
\end{figure*}

\begin{figure*}
    \centering
    \includegraphics[width=0.98\textwidth]{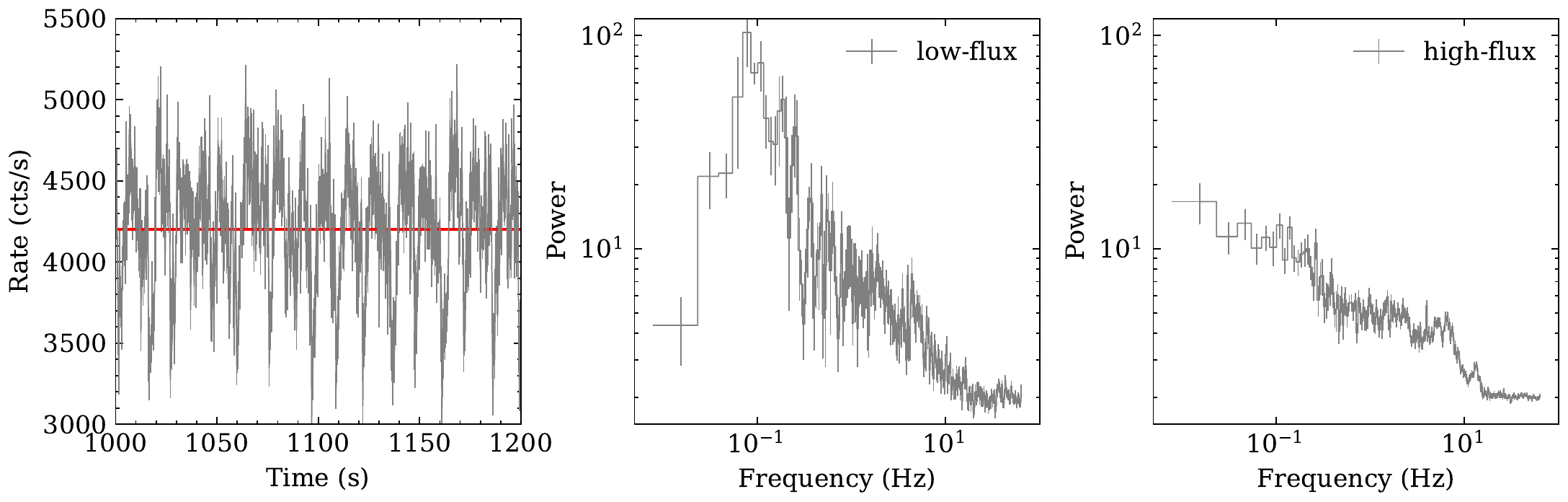}
    \caption{Left panel: the representative lightcurve. Middle panel: the PDS of low-flux. Right panel: the PDS of high-flux. The low-flux and high flux is divided based on the red line of left panel.}
    \label{Fig:Fig12}
\end{figure*}

\begin{figure}
    \centering
    \includegraphics[width=0.4\textwidth]{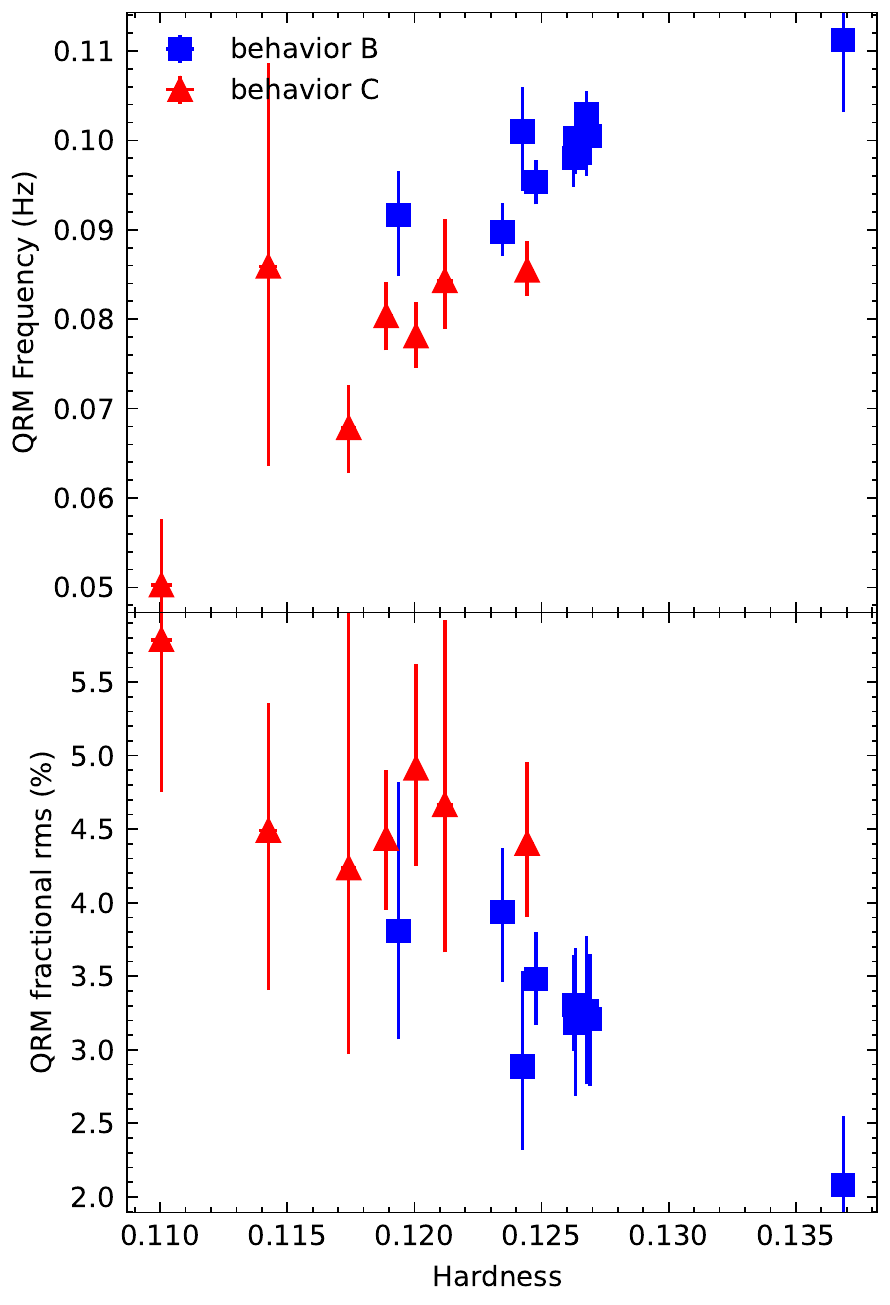}
    \caption{Top panel: the centroid frequency of the mHz QRM versus hardness. Bottom panel: the fractional rms of the mHz QRM versus hardness ratio. Hardness ratio is defined as the count rate ratio between 1.94--11.91 keV and 11.91--30.02 keV.}
    \label{Fig:Fig13}
\end{figure}

\begin{figure}
    \centering
    \includegraphics[width=0.4\textwidth]{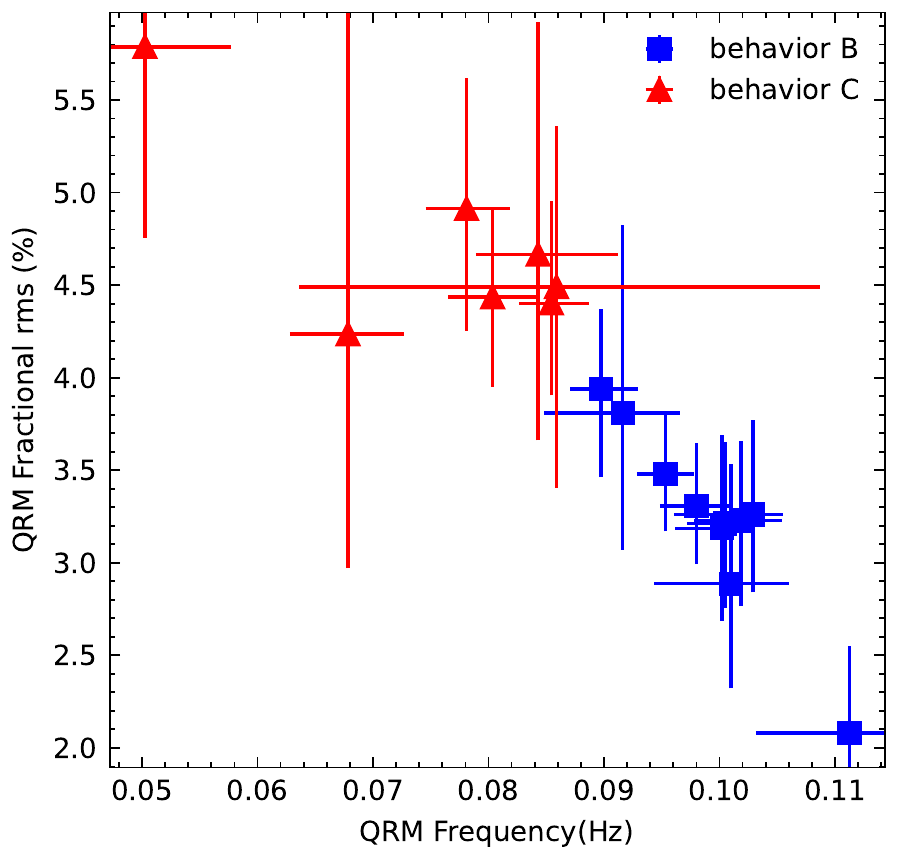}
    \caption{The centroid frequency versus the fractional rms of mHz QRM.}
    \label{Fig:Fig14}
\end{figure}

\begin{figure*}
    \centering
    \includegraphics[width=0.98\textwidth]{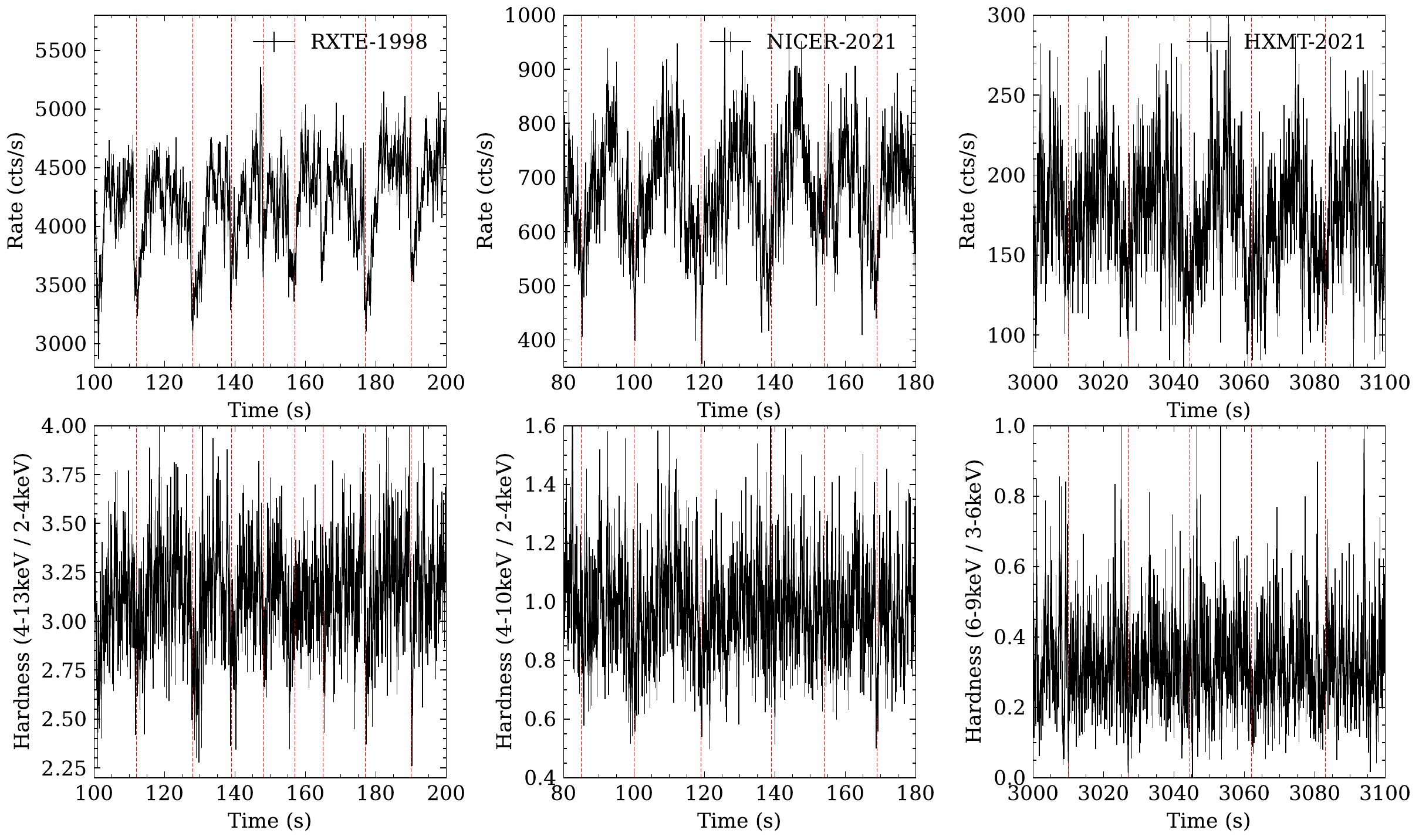}
    \caption{Representative light curves and hardness ratio of RXTE, NICER, and Insight-HXMT. The red lines mark the dip of light curves}
    \label{Fig:Fig15}
\end{figure*}

\subsection{Energy spectra}
 We first try \textsc{Tbabs*(diskbb+cutoffpl)} model in our spectral fittings, but we obtain a large $\chi^2$ (generally > 1.5). To obtain a moderate $\chi^2$, we add the {\sc relxill} model in our spectral fittings \citep{2014ApJ...782...76G}. The contribution of the disc component is not significant, possibly due to the high nH absorption and the energy range of PCA, and so even if the disc component is removed, we can still obtain a moderate reduced $\chi^2$ (for most observations, $\chi^2$ <= 1.20). We set the cut-off energy ($E_{\rm cut}$) of the {\sc cutoffpl} model free to fit, however, for the less constrained cut-off energy ($E_{\rm cut}$), we fix it in 300 keV \citep{2010ApJ...718..695G}. For the {\sc relxill} model, we fix the abundance of iron to 5 based on our preliminary fits. As for the spin and inclination, we adopt \citet{king_disk_2014} fitting results, $a_\ast$ = 0.985, i = 64 deg. To only calculate the reflection emission, the reflection fraction is taken as -1 \citep{2014ApJ...782...76G}. We set the power-law emissivity as $r^{-3}$ \citep{1989MNRAS.238..729F}. The $ \Gamma$ and $E_{\rm cut}$ are tied to the corresponding parameters in {\sc cutoffpl}. Other parameters in model {\sc cutoffpl} are free in spectral fitting. The fitting results are listed in table \ref{table:table2}. $\Gamma$, $ R_{\rm in}$ and log $\xi$ show no obvious evolutionary trends. 

In Fig.~\ref{Fig:Fig11}, we show the frequency and the fractional rms of mHz QRM as a function of a certain spectral component flux ratio. We use cflux command (a convolution model) to calculate the flux in the energy range of 2.5--30 keV for different models. From the panel c of Fig. \ref{Fig:Fig11}, we can see that the QRM fractional rms of behavior B has a positive correlation with $\rm F_{\rm powerlaw}$ / $\rm F_{\rm total}$ and an anti-correlation with $\rm F_{\rm reflection}$ / $\rm F_{\rm total}$, where $\rm F_{\rm total}$ = $\rm F_{\rm powerlaw}$+$\rm F_{\rm reflection}$. However, the frequency and rms of behavior C have no obvious correlation with $\rm F_{\rm powerlaw}$ / $\rm F_{\rm total}$ and $\rm F_{\rm reflection}$ / $\rm F_{\rm total}$.

\section{Discussion}
We have studied the spectral timing properties of the mHz QRM shown in 4U 1630--47 during the outburst in 1998. The main results are summarized as follows:

(1) As shown in Fig.~\ref{Fig:Fig3}, the mHz QRM only appears in a specific flux range ($\sim$ 1.40 $\times$ $10^{-8}$ erg $~\rm s^{-1}$  $~\rm cm^{-2}$, 3--20 keV). When the flux suddenly increases, the mHz QRM disappears.

(2) The power density spectrum shows two different behaviors in the mHz QRM time zones. For behavior B and C, as shown in Figs.~\ref{Fig:Fig4} and~\ref{Fig:Fig9}, the properties of the mHz QRM are basically similar. The energy dependence of the fractional rms on the mHz QRM shows a hump in the range of 5--7 keV, then increases with photon energy. Below $\sim$ 13 keV, the phase-lag of mHz QRM exhibits a positive value. Above $\sim$ 13 keV, it turns into a soft lag and the soft lag increases with photon energy.

(3) As shown in Fig.~\ref{Fig:Fig2}, the hardness of behavior B is higher than behavior C. For behavior B, its PDS has two LFQPOs features, separately at $\sim$ 5 Hz \& $\sim$ 13 Hz. For behavior C, its PDS has three LFQPOs features $\sim$ 4 Hz, $\sim$ 7 Hz, $\sim$ 13 Hz, respectively. As pointed out in Fig.~\ref{Fig:Fig5}, we can see that the LFQPO frequency ($\sim$ 5 Hz) of behavior B increases with photon energy; the LFQPO frequency ($\sim$ 4 Hz) of behavior C is roughly constant at different energy bands; the LFQPO frequency ($\sim$ 7 Hz) of behavior C decreases with photon energy; the LFQPO ($\sim$ 5 Hz) fractional rms of behavior B increases with energy below $\sim$ 10keV, then decreases with energy; the LFQPO fractional rms of behavior C ($\sim$ 4 Hz) seems to be constant at different energy bands, and the LFQPO ($\sim$ 7Hz) fractional rms of behavior C increases with photon energy. As shown in Fig.~\ref{Fig:Fig9} bottom panel, the phase-lag energy dependence of the LFQPO ($\sim$ 5 Hz) of behavior B is similar with the LFQPO($\sim$ 4Hz) of behavior C, and is opposite with the LFQPO ($\sim$ 7Hz) of behavior C.  

(4) As shown in Table~\ref{table:table2}, $\Gamma$, $ R_{\rm in}$ and log $\xi$ have no significant change.

(5) In Fig.~\ref{Fig:Fig11}, the mHz QRM frequency of behavior B has a positive correlation with $\rm F_{\rm reflection}$ / $\rm F_{\rm total}$ and an anti-correlation with $\rm F_{\rm powerlaw}$ / $\rm F_{\rm total}$. The QRM fractional rms of behavior B has a positive correlation with $\rm F_{\rm powerlaw}$ / $\rm F_{\rm total}$ and an anti-correlation with $\rm F_{\rm reflection}$ / $\rm F_{\rm total}$. But both the frequency and rms of behavior C have no obvious correlation with $\rm F_{\rm powerlaw}$ / $\rm F_{\rm total}$ and $\rm F_{\rm reflection}$ / $\rm F_{\rm total}$.

\subsection{Comparison with other low frequency modulation phenomena}

\subsubsection{Comparison with 2021 outburst in 4U 1630-47}
As mentioned above, 4U 1630--47 exhibits the mHz QRM phenomenon during its outburst in 2021 \citep{2022ApJ...937...33Y}. In this outburst, the frequency of the mHz QRM is about $\sim$ 0.06 Hz. For the 1998 outburst, the frequency of the mHz QRM is in the range of $\sim$ 0.05-0.1 Hz \citep{dieters_timing_2000, trudolyubov_rxte_2001}. However, in the 1998 outburst, the mHz QRM and LFQPO are both detected simultaneously, whereas in the 2021 outburst, the mHz QRM appears after LFQPO disappears \citep{2022ApJ...937...33Y}. The mHz QRM appears in a very close flux range ($\sim$ 1.40 $\times$ $10^{-8}$ erg $~\rm s^{-1}$  $~\rm cm^{-2}$, 3--20 keV) for both outbursts. For the relationship between the centroid frequency and photon energy of the mHz QRM, it remains almost constant in 1998, while it shows a weak positive correlation in 2021 with very large errorbars. However, considering the same energy range covered by the instruments, the relationship between the two outbursts is similar. For the relationship between the fractional rms and energy of the mHz QRM, the trends in 1998 and 2021 are also similar, except for a hump in the range of 5--7 keV, possibly due to the lower sensitivity of the LE detector of Insight-HXMT. This trend is different from that of LFQPOs exhibited by other sources, for instance: H 1743-322 \citep{2013MNRAS.433..412L}, XTE J1550-564 \citep{2013MNRAS.428.1704L}, MAXI J1535-571 \citep{2018ApJ...866..122H}, GRS 1915+105 \citep{2016ApJ...833...27Y}, and MAXI J1653-479 \citep{2022RAA....22k5002F}. The fractional rms of LFQPOs seen in these sources first increases with photon energy below 10 keV, and gradually saturates after 10 keV. Besides, the phase-lag energy dependence of the mHz QRM is also similar in 1998 and 2021. Below $\sim$ 13 keV, it is a slight hard lag, but when the energy exceeds $\sim$ 13 keV, it turns into a soft lag, and the soft lag value increases with energy. The maximum phase-lag reached $\sim$ 2 rad ($\sim$ 3 seconds), indicating that it could not be due to compton scattering or reflection processes \citep{uttley_causal_2011}. Many properties exhibited by 4U 1630-47 in the two outbursts in 2021 and 1998 are very similar, so we can reasonably speculate that the mHz QRM of the two outbursts may harbor the same physical origin. For the LFQPO simultaneous with the mHz QRM during the 1998 outburst, it may be produced from a process independent from mHz QRMs.

\subsubsection{Comparison with the mHz QRM in H 1743-322}

\citet{altamirano_low-frequency_2012} also reported the phenomenon of the mHz QRM and LFQPOs simultaneously detected in H 1743--322. They observed the mHz QRM at the similar hardness and intensity ($\leq$ 4.0 $\times$ $10^{-9}$ erg $~\rm s^{-1}$ $~\rm cm^{-2}$, 2--20 keV) during the 2010 \& 2011 outbursts, and suggested that the mHz QRM may have a dependence on accretion state. They found that the frequency of mHz QRM remains almost constant, while the LFQPOs frequency varies by a factor of $\sim$ 2. Therefore, they concluded the mechanisms that produce the mHz QRMs and LFQPOs are independent. The mHz QRMs of H 1743--322 were observed in the rising state of the 2010 \& 2011 outbursts with a high Q factor. But the mHz QRMs of 4U 1630--47 were observed in the intermediate state. The fractional rms energy dependence of the mHz QRMs in H 1743--322 is different from 4U 1630--47. For the 2010 outburst, the fractional rms of mHz QRMs is roughly constant in different energy bands. For the 2011 outburst in H 1743-322, the fractional rms firstly increases with energy, then decreases with energy. But the increasing trend is not significant ( $\leq$ 1$\%$). They speculate the mHz QRMs may be related to the radio jet. However, due to the difference in the properties of the mHz QRM between H 1743--322 and 4U 1630--47, we can not confirm that the mHz QRM signals of the two sources are from the same origin.

\subsection{Possible models to explain the mHz QRM}

\subsubsection{Geometric precession toy model}
In Fig.~\ref{Fig:Fig4}, we can clearly see that QRM fractional rms increases with the photon energy above $\sim$ 10 keV. As mentioned above, the 1998 and 2021 outbursts of 4U 1630--47 show many similar properties. Therefore, if we consider the mHz QRM of the two outbursts as being generated by similar mechanisms, then they should behave similarly at higher energy bands. As \citet{2022ApJ...937...33Y} showed in their Fig. 5, the fractional rms increases with the energy up to 60-100 keV. This result implies that the mHz QRM may origin from the hot corona region. Meanwhile, we also notice that LFQPOs and mHz QRM are both detected during 1998 outburst. In general, the canonical LFQPO is thought to originate from the corona or jet base \citep{ingram_low-frequency_2009, ingram_review_2019, ma_discovery_2021,2021ApJ...919...92B}. The frequency of QPO signals are strongly dependent on the emission radius of the corona/jet base. \citet{2016MNRAS.458.3655V} proposed that a geometric precession toy model could be an interesting interpretation to explain the simultaneous detection of LFQPOs and mHz QRMs. As shown in Fig. \ref{Fig:Fig13}, there is a positive correlation between the centroid frequency and hardness of the mHz QRM. This correlation is different from type-C QPOs, which usually has an inverse relationship between centroid frequency and hardness. There is an inverse correlation between the fractional RMS of mHz and hardness, which is also the opposite of the type-C QPO. In Fig.~ \ref{Fig:Fig14}, we show the fractional rms versus centroid frequency of the mHz QRM. It is very similar to the trend of type-C QPOs. Therefore, the mHz QRM in this outburst may originate from a precession process with a larger radius similar to type-C QPO's. In this scenario, LFQPOs and mHz QRMs are generated by different precession torus located at different radii, the corresponding precession radius for the inner torus is $\sim$ 11 $R_g$, and the outer torus is $\sim$ 40 $R_g$ by considering a 10 $\rm M_\odot$ \citep{seifina_black_2014} and  $a_\ast$  $\thicksim$ 0.985 \citep{king_disk_2014}. This requires a large truncation radius. But from the spectral fitting result shown in Table~\ref{table:table2}, we can see that the radius of inner disc is in the range of $\sim$ 2--5 $R_{\rm g}$. Therefore, this result is not consistent with the requirements of truncation radius. In Fig.~\ref{Fig:Fig12}, we divide the light curve into a high-flux interval and a low-flux interval. The mHz QRM is detected in low-flux interval while the LFQPOs are detected in high-flux interval. If the light curve is modulated by two independent Lense-Thirring precessions, the LFQPOs and mHz QRMs should be detected both in low-flux and high-flux intervals. So, this model cannot account for this phenomenon.

\subsubsection{Instability models}
As mentioned above, \citet{2022ApJ...937...33Y} has reported a mHz QRM phenomenon in the 2021 outburst of 4U 1630--47. Based on evidence showing that the fractional rms is positively correlated with energy in higher energy bands, they constrain the physical origin of this mHz QRM in the hot corona region, and they hypothesize that the mHz QRM signals could be interpreted by an unknown instability in the corona. An unknown instability with strict accretion rate dependence occurs in the expanding corona, which then leads to this quasi-regular flux modulation. From Fig.~\ref{Fig:Fig4}, we can also see that the fractional rms has a positive correlation with energy in higher energy bands. And in Fig.~\ref{Fig:Fig11} panel c: the QRM fractional rms has a positive correlation with $\rm F_{\rm powerlaw}$ / $\rm F_{\rm total}$. Based on the previous discussion, we speculate that the origin of the mHz QRM  of the two outbursts of 4U 1630--47 is similar. Therefore, we can also speculate the origin of mHz QRM observed in 1998 outburst is also in the hot corona region. However, for the 1998 outburst of 4U 1630--47, LFQPOs and the mHz QRMs are  observed simultaneously. In general, the canonical LFQPO is thought to originate from the corona or jet base \citep{ingram_low-frequency_2009, ingram_review_2019, ma_discovery_2021,2021ApJ...919...92B}. If we consider the mHz QRM produced from the instability of the corona, the expanding corona would change the frequency of LFQPOs. But the frequency of LFQPOs show no significant change when mHz QRMs appear as shown in Fig.~\ref{Fig:Fig3}. It is obviously hard to simultaneously harbor two oscillations of completely different frequencies produced from the instability of a corona or jet base which usually has a limited size. However, if we consider the mHz QRM and LFQPOs as being produced from two different sources (corona/jet), it would be an interesting interpretation. One possibility in this scenario is that LFQPOs are generated by the Lense-Thirring precession of the jet above the black hole, and the mHz QRM is produced from the instability of the corona in the inner region. It should be noted, however, that our assumption is that the jet precesses with the corona below as a whole \citep{2018MNRAS.474L..81L, 2023arXiv230300481M}, and it remains uncertain how the instability of the corona influences the precessing jet. If we consider the opposite scenario where the mHz QRM is generated from the Lense-Thirring precession of the jet, and the LFQPOs are generated from the instability of the corona,  a large truncation radius is required ($\sim$ 40 $R_g$). The appearance of the mHz QRM when the source is in intermediate states indicates that the truncation radius of the disc should not be very large. Additionally, the value of $R_{\rm in}$ of the energy spectral fitting result suggests that the inner radius ranges from $\sim$ 2--5 $R_{\rm g}$. Therefore, the mHz QRM is not likely to be generated from the Lense-Thirring precession of the jet.

However, Lense-Thirring precession is not the only plausible explanation for LFQPOs. According to \citet{2022A&A...662A.118M}, oscillations can be produced from the interaction between the hot Comptonizing corona and the cold accretion disc. The frequencies of these oscillations exhibit similarities to frequencies of type-C QPOs observed in Black-hole X-ray binaries. Therefore, they argued that the frequencies of type-C QPOs are the natural frequencies in a dynamical disc-corona system. In their model, the frequency of oscillation depends on the accretion rate, the size of the corona, and the ratio of the energy densities of reprocessed soft photons and hard photons. As shown in Fig.~\ref{Fig:Fig3}, when the mHz QRM exists, the flux remains at a stable level, so the accretion rate is approximately constant. This indicates that the disc-corona system may reach a balance, so the frequency of mHz QRMs do not show significant change. As we mentioned before, the mHz QRM only appears in a certain accretion rate, corresponding to $L_{2-100 \rm \ keV} \sim 0.16 \ L_{\rm Edd}$ (assuming a distance of $D=10 \ {\rm kpc}$). For typical parameter values of BHXRBs, at this level of accretion, the natural frequency range of the disc-corona system is between approximately 1 Hz and tens of Hz (see fig. 5 in \citet{2022A&A...662A.118M}). The predicted oscillation frequency of this model is higher than the frequency of the mHz QRM observed in 4U 1630--47. Therefore, the mHz QRM is not likely to be produced by this process. However, the LFQPOs may be produced from this process. Based on this model, one possible scenario is that the LFQPOs are generated from the interaction between the corona and the disc, and the mHz QRM may be generated from the instability of jet.

Variable Comptonization models are also one popular explanation for LFQPOs. \citet{2022NatAs...6..577M} proposed a coronal-jet transition scenario based on their analysis of over 1800 observations of GRS 1915+105 during the RXTE era, which accounts for the evolution of LFQPO frequency and phase-lag. They observed that LFQPOs below 2 Hz exhibit negative lag, while those above 2 Hz exhibit positive lag. In their proposed model, the reflection process dominates the lag for the former, while the Compton scattering process dominates the lag for the latter. In Fig.~\ref{Fig:Fig9}, we show the phase-lag of the mHz QRM and LFQPOs as a function of energy. The time-lag of the mHz QRM is on the order of seconds, so we can not use the Compton scattering or reflection processes to explain time-lag behavior of the mHz QRM, because the magnitude of Compton scattering or light time traveling time lags is on the order of milliseconds. However, the time-lag for LFQPOs is at the millisecond level ($\sim$10 ms), which can be produced by the above mentioned process, indicating that mHz QRMs and LFQPOs may not share the same origin. Moreover, an interesting aspect of this model is that during the inward movement of the accretion disc, the jet can transform into a corona. Therefore, the corona and jet can coexist when the accretion rate is at a certain level. In this scenario, for 4U 1630--47, LFQPOs may arise from the corona, while mHz QRMs could originate from the instability of the jet. \citet{2022MNRAS.515.2099B} proposed a time-dependent Comptonization model: vKompth, in which the diskbb spectrum is taken as the soft photons source. Their model describes the QPO as a small fluctuation of the spectrum around the time-averaged spectrum. The small oscillation of the spectrum around the time-averaged spectrum in their model can be generated by changes in any physical quantity incorporated in the Kompaneets equation. Specifically, they hypothesize that variations in the coronal temperature are responsible for the fluctuations. This model has been successfully applied to the type-B QPO of MAXI J1438--630 and the type-C QPO of GRS 1915+105. Their model shows that the rms amplitude increases from about 1$\%$ to 10$\%$ with energy, and flattens out above 10 keV. Although this trend is different from the one shown in Fig.~\ref{Fig:Fig4}, it is roughly consistent with the trends in Fig.~\ref{Fig:Fig5} (except for Behavior C $\sim$ 4 Hz). Therefore, the energy dependence of rms for LFQPOs can well explained by this model. \citet{2022MNRAS.513.4196G} applied the variable-Comptonization models to analyze the type-C QPOs of GRS 1915+105. The energy-depentent phase-lag spectra and fractional rms spectra can be well-fitted by the spectral-timing Comptonization model. And for the $\sim$ 4Hz QPO observed in GRS 1915+105, it exhibits a soft lag behavior, similar to Behavior B $\sim$ 5 Hz and Behavior C $\sim$ 4 Hz) in Fig.~\ref{Fig:Fig9}. Therefore, the phase-lag behaviors of LFQPOs can be also well explained by variable-Comptonization model. 

For lags on the order of seconds like mHz QRM, viscous fluctuations would be a possible reason. \citet{1997MNRAS.292..679L} considered an $\alpha$ parameter at different radius and found that the $\alpha$ fluctuations characteristic time-scale is of the order of the local viscous time-scale. \citet{2016MNRAS.457.2999M} proposed a delayed response of the inner disc radius to the accretion rate. For their model, the power-law component is immediately responding and the response of the inner radius to the accretion rate has a delay time up to several seconds. The phase-lag is a hard lag in the low energy bands and turn to a soft lag in the high energy bands. This is similar to the result shown in Fig.~\ref{Fig:Fig9}. However, they found the energy dependence of the fractional rms increases with energy below $\sim$ 10 keV, then decreases with energy above $\sim$ 10 keV.  This result is not consistent with Fig.~ \ref{Fig:Fig4}. Therefore, \citet{2016MNRAS.457.2999M} model can not explain the relationship of Fig.~\ref{Fig:Fig4}.

\subsubsection{Periodic occlusion}

In Fig.~\ref{Fig:Fig15}, we show the representative light curve and hardness of RXTE (1998), NICER (2021) and \hxmt\ (2021), respectively. For the 1998 outburst, we can see that sharp dips exist in the light curve. The dips are usually triangular in shape, dropping rapidly ($\sim$ 1s) and then rising ($\sim$ 3-5 s) \citep{dieters_timing_2000}. For dips in light curve, occlusion by certain material is a possible reason. If we consider that the quasi-regular flux modulation (QRM) is generated by the occlusion of the material block on the outer disc, then the radius corresponding to the Kepler frequency of the mHz QRM is approximately 1.1 $\times 10^{9}$ cm ($\sim$ 733 $R_{\rm g}$). With an inclination of 64 deg, the size of the matter block should be about $358 R_{\rm g}$ to form an occlusion calculated from H = R / $\tan(i)$, H is the height of the matter block, R is the radius of the matter block, i is the inclination. In this case, the material block is too large for a standard disc \citep{shakura_black_1973}. However, if we consider that the inner disc is in a warped state, the size requirement for the material block would be smaller.

If the equatorial plane of the accretion disc and the black hole are not in the same plane, the spinning black hole will cause non-equatorial orbits to precess. Differential precession may cause a warping in an accretion disc. If the disc is relatively thin ( scale height $H/R < \alpha$, $\alpha$ is the dimensionless viscosity of the disc),  warps will be propagated by viscous diffusion. Therefore, it might be possible that the inner disc is aligned with the black hole's axis of rotation while the outer disc remains titled \citep{1975ApJ...195L..65B}. \citet{2012MNRAS.421.1201N} found that non-linear fluid effects can promote breaking the disc into distinct planes by simulating the viscous evolution of an accretion disc around a spinning black hole. \citet{2021MNRAS.507..983L} found that the Lense-Thirring effect of a rotating black hole overpowers the accretion disc. The disc will be torn into a rapidly precessing inner sub-disc surrounded by a slowly precessing outer sub-disc. The radius of warped occurs can be estimated by the following formula \citep{2014MNRAS.437.3994N}: 

$$R_{\rm warped} \thicksim 2^{4/3}  \left(\frac{a_\ast}{\alpha}\right)^{2/3}  \left(\frac{R}{H}\right)^{4/3} R_{\rm g}$$

The warped transition radius is dependent on a number of detailed assumptions, and it still under debate \citep{2001ApJ...553..955F, 2013Sci...339...49M, 2015MNRAS.449.1251D}. Here we consider $a_\ast$ = 0.998, $\alpha$ = 0.1 \citep{2007MNRAS.376.1740K}, H/R = 0.02 \citep{shakura_black_1973, 2012MNRAS.421.1201N}. $R_{\rm warped}$ is $\sim$ 2000 $R_{\rm g}$. Therefore, the warped disc may form an occlusion at the Kepler radius of mHz QRM by considering a large observation inclination.  Since the warped disc is optically thick, the hard photon will be reprocessed, and the hardness ratio may present a dip feature as shown in \ref{Fig:Fig15}. When the flux suddenly increases, the warped disk may be unstable, so it can not form periodic occlusions, leading to the disappearance of mHz QRMs; when the flux returns to the former range, the disc may return to a warped state, so the mHz QRM is detected again as shown in Fig.~\ref{Fig:Fig2}. In this case, the mechanisms of LFQPOs and mHz QRM are independent processes. The former is generated by the precession of the corona or jet, while the latter is generated by the occlusion of the warped disc. The more inner regions will be more heavily occluded due to the geometry effects, so the fractional rms will be higher. The frequency of the mHz QRM is only related to the radius of the material block, so the centroid frequency of the mHz QRM for different energy bands remains roughly constant. If occlusion was observed, the system should have a high inclination. As mentioned in section 1, 4U 1630--47 has a high inclination (60-70 deg). Besides 4U 1630--47, several sources with the mHz QRM observation also have a high inclination, for instance: H 1743--322 \citep{altamirano_low-frequency_2012}, which suggests that the appearance of the mHz QRM may be related to the inclination.

\section{Conclusion}

We present here a detailed timing and spectral study on the mHz QRM phenomenon of 4U 1630--47 during its 1998 outburst. The QRM in the 1998 outburst is flux dependent, similar to that in the 2021 outburst, and the energy dependence of the timing properties are also similar to the 2021 outburst. But the QRM during the 1998 outburst is detected simultaneously with LFQPOs.  For this mHz QRM, we have proposed some ideas to interpret. We need more observations to have a better understanding of the mHz QRM phenomena.

\section*{Acknowledgements}

We greatly appreciate for the anonymous referee’s helpful comments. This research has made use of data obtained from the High Energy Astrophysics Science Archive Research Center(HEASARC), provided by NASA’s Goddard Space Flight Center.
This work is supported by the Natural Science Foundation of China under grant No. 11873035 and the National Key R\&D Program of China (2021YFA0718500). We acknowledges funding support from the National Natural Science Foundation of China (NSFC) under grant No. 12122306, the CAS Pioneer Hundred Talent Program Y8291130K2 and the Scientific and technological innovation project of IHEP Y7515570U1.
Erlin Qiao is  supported by the National Natural Science Foundation of China (grants 12173048) , NAOC Nebula Talents Program.

%%%%%%%%%%%%%%%%%%%%%%%%%%%%%%%%%%%%%%%%%%%%%%%%%%
\section*{Data Availability}

The observational data of RXTE used in this article are available from the HEASARC archive https://heasarc.gsfc.nasa.gov. 

The raw data of HXMT uesd in this article are available at http://hxmten.ihep.ac.cn/.

%%%%%%%%%%%%%%%%%%%% REFERENCES %%%%%%%%%%%%%%%%%%

% The best way to enter references is to use BibTeX:

\bibliographystyle{mnras}
\bibliography{refer.bib/ref.bib} % if your bibtex file is called example.bib

% Alternatively you could enter them by hand, like this:
% This method is tedious and prone to error if you have lots of references
%\begin{thebibliography}{99}
%\bibitem[\protect\citeauthoryear{Author}{2012}]{Author2012}
%Author A.~N., 2013, Journal of Improbable Astronomy, 1, 1
%\bibitem[\protect\citeauthoryear{Others}{2013}]{Others2013}
%Others S., 2012, Journal of Interesting Stuff, 17, 198
%\end{thebibliography}

%%%%%%%%%%%%%%%%%%%%%%%%%%%%%%%%%%%%%%%%%%%%%%%%%%

%%%%%%%%%%%%%%%%% APPENDICES %%%%%%%%%%%%%%%%%%%%%

\appendix

\section{Some extra material}
\begin{table*}
    \caption{The list of RXTE/PCA observations in 1998 outburst of 4U 1630-47 used in the analysis. Centroid Frequency and Fractional RMS are measured for QRM/QPO component of 2--13 keV. Q factor is defined as the centroid frequency/FWHM. a: the frequency is lower significant than other observations, so we exclude this observation in statistical analysis below.}
    \scalebox{0.85}{
        \begin{tabular}{llllllll}
        \hline
        \multicolumn{1}{|l|}{$\#$} &
        \multicolumn{1}{l|}{ObsID} &
        \multicolumn{1}{l|}{ MJD } &
        \multicolumn{1}{l|}{ Exposure (s)} &
        \multicolumn{1}{l|}{ Behavior}  &
        \multicolumn{1}{l|}{ Centroid Frequency (Hz)} &
        \multicolumn{1}{l|}{ Fractional rms ($\%$)} &
        \multicolumn{1}{l|}{ Q factor} \\ \hline
    1  & 30188-02-04-00 & 50856.97  & 2096   & A & $4.78^{+0.04}_{-0.04}$   & $4.8^{+0.3}_{-0.3}$  & $4.2^{+0.6}_{-0.5}$   \\
      &    &     &  & & $8.0^{+0.2}_{-0.2}$   & $3.5^{+0.3}_{-0.3}$  & $3.2^{+0.6}_{-0.5}$   \\
    2  & 30178-01-05-00 & 50857.12  & 1497  & B & $0.111^{+0.009}_{-0.007}$ & $2.1^{+0.5}_{-0.5}$ & $5.0^{+3.9}_{-4.9}$   \\
       &    &     &  & & $5.34^{+0.08}_{-0.10}$   & $2.5^{+0.8}_{-0.5}$  & $7.0^{+4.3}_{-2.7}$   \\
       &    &     &  & & $7.7^{+0.3}_{-0.2}$    & $1.5^{+0.4}_{-0.4}$  & $7.7^{+0.3}_{-0.2}$   \\
    3  & 30188-02-05-00 & 50857.71  & 2569  &  C & $0.068^{+0.005}_{-0.005}$ & $4.2^{+2.1}_{-1.3}$ & $2.3^{+2.9}_{-2.1}$  \\
     &    &     &  & & $4.93^{+0.07}_{-0.08}$   & $1.5^{+0.2}_{-0.2}$  & $12.3^{+0.2}_{-0.2}$   \\
       &    &     &  & & $7.0^{+0.1}_{-0.1}$    & $3.6^{+0.2}_{-0.2}$  & $4.7^{+0.6}_{-0.6}$   \\
    &    &     &  & & $13.8^{+0.2}_{-0.2}$   & $1.8^{+0.2}_{-0.2}$  & $7.8^{+2.8}_{-2.2}$   \\
    4  & 30178-02-02-00 & 50857.79  & 7865  &  C & $0.078^{+0.004}_{-0.004}$ & $4.9^{+0.7}_{-0.7}$ & $2.3^{+2.9}_{-2.1}$  \\
       &    &   &  &  & $4.9^{+0.1}_{-0.1}$   & $1.4^{+0.2}_{-0.2}$  & $12.3^{+0.2}_{-0.2}$   \\
       &    &   &  &  & $7.0^{+0.1}_{-0.1}$   & $3.7^{+0.4}_{-0.4}$  & $3.6^{+0.6}_{-0.6}$   \\
       &    &   &  &  & $13.9^{+0.2}_{-0.2}$   & $1.8^{+0.2}_{-0.2}$  & $9.6^{+3.7}_{-2.6}$   \\
    5  & 30188-02-06-00 & 50858.05 & 2433  &  B & $0.101^{+0.002}_{-0.003}$ & $3.2^{+0.4}_{-0.4}$ & $6.9^{+3.9}_{-4.3}$  \\
        &    &   &  &  & $5.5^{+0.1}_{-0.1}$   & $4.2^{+0.7}_{-0.5}$  & $2.3^{+0.7}_{-0.5}$   \\
       &    &   &  &  & $13.5^{+0.2}_{-0.3}$   & $4.2^{+0.7}_{-0.5}$  & $2.3^{+0.7}_{-0.5}$   \\
    6  & 30178-01-06-00 & 50858.70  & 1765  &  B & $0.103^{+0.003}_{-0.007}$ & $3.3^{+0.5}_{-0.4}$ & $5.9^{+6.6}_{-3.0}$ \\
    &    &   &  &  & $5.4^{+0.2}_{-0.2}$   & $4.9^{+0.7}_{-0.7}$  & $2.0^{+0.6}_{-0.5}$   \\
    &    &   &  &  & $12.5^{+0.5}_{-0.5}$   & $1.9^{+0.4}_{-0.4}$  & $4.8^{+2.6}_{-1.9}$   \\
    7  & 30188-02-07-00 & 50858.72  & 2036  &  B & $0.102^{+0.004}_{-0.004}$ & $3.2^{+0.4}_{-0.5}$ & $4.3^{+2.3}_{-2.0}$  \\
    &    &   &  &  & $5.4^{+0.1}_{-0.1}$   & $4.4^{+0.7}_{-0.5}$  & $2.3^{+0.7}_{-0.5}$   \\
    &    &   &  &  & $13.3^{+0.3}_{-0.3}$   & $1.9^{+0.4}_{-0.3}$  & $7.3^{+1.9}_{-1.5}$   \\
    8  & 30178-02-02-01 & 50858.77  & 9229  &  B & $0.095^{+0.002}_{-0.002}$ & $3.7^{+0.3}_{-0.3}$ & $3.5^{+0.9}_{-0.8}$  \\
    &    &   &  &  & $5.4^{+0.1}_{-0.1}$   & $5.9^{+0.2}_{-0.2}$  & $1.3^{+0.1}_{-0.1}$   \\
    &    &   &  &  & $13.2^{+0.3}_{-0.3}$   & $5.9^{+0.2}_{-0.2}$  & $1.3^{+0.1}_{-0.1}$   \\
    9  & 30178-01-07-00 & 50859.84  & 1426  &  C & $0.086^{+0.002}_{-0.002}$ & ${4.5}^{+0.9}_{-1.1}$ &   $1.0^{+0.6}_{-0.5}$   \\
    &    &   &  &  & $5.2^{+0.1}_{-0.1}$   & $4.8^{+0.7}_{-0.6}$  & $1.8^{+0.4}_{-0.3}$   \\
    &    &   &  &  & $6.91^{+0.05}_{-0.05}$   & $3.4^{+0.2}_{-0.2}$  & $6.3^{+0.8}_{-0.8}$   \\
    10  & 30188-02-08-00 & 50860.12  & 2067 & C & $0.028^{+0.014}_{-0.020}$ & $6.5^{+0.3}_{-0.3}$ & $0.2^{+0.1}_{-0.1}$ \\  
    &   &   &  &   & $4.3^{+0.2}_{-0.2}$   & $1.6^{+0.4}_{-0.3}$  & $5.0^{+3.6}_{-2.1}$   \\
    &    &  &   &  & $6.88^{+0.06}_{-0.06}$   & $2.9^{+0.2}_{-0.2}$  & $7.4^{+1.1}_{-1.0}$   \\
    11 & 30188-02-09-00 & 50860.56  & 1835  &  B & $0.092^{+0.005}_{-0.007}$ & $3.8^{+1.0}_{-0.7}$ & $3.0^{+2.9}_{-1.6}$  \\
    &    &   &  &  & $5.19^{+0.08}_{-0.08}$   & $1.9^{+0.2}_{-0.2}$  & $10.4^{+0.2}_{-0.2}$   \\
    &    &   &  &  & $7.0^{+0.3}_{-0.2}$   & $2.6^{+0.2}_{-0.2}$  & $7.0^{+0.3}_{-0.2}$   \\
    &    &   &  &  & $13.4^{+0.5}_{-0.6}$   & $1.0^{+0.4}_{-0.3}$  & $11.2^{+14.6}_{-8.1}$   \\
    12 & 30178-01-08-00 & 50860.72  & 1453  &  B & $0.101^{+0.005}_{-0.007}$ & $2.9^{+0.9}_{-0.6}$ & $4.5^{+3.6}_{-3.1}$  \\ 
    &    &   &  &  & $5.3^{+0.1}_{-0.1}$   & $4.6^{+0.4}_{-0.4}$  & $2.4^{+0.6}_{-0.4}$   \\
    &    &   &  &  & $13.2^{+0.4}_{-0.4}$   & $1.6^{+0.4}_{-0.3}$  & $7.3^{+5.9}_{-3.2}$   \\
    13 & 30178-01-09-00 & 50861.69  & 1719  &  B & $0.100^{+0.002}_{-0.004}$ & $3.2^{+0.5}_{-0.5}$ & $8.0^{+6.2}_{-7.7}$  \\
    &    &   &  &  & $5.3^{+0.1}_{-0.1}$   & $3.2^{+0.4}_{-0.3}$  & $4.3^{+1.4}_{-1.0}$   \\
    &    &   &  &  & $13.9^{+0.2}_{-0.2}$   & $1.3^{+0.4}_{-0.3}$  & $15.4^{+16.5}_{-9.5}$   \\
    14 & 30188-02-10-00 & 50861.72  & 2027  &  B & $0.090^{+0.003}_{-0.003}$ & $3.9^{+0.4}_{-0.5}$ & $5.7^{+2.3}_{-2.7}$  \\
    &    &   &  &  & $5.0^{+0.3}_{-0.3}$   & $2.6^{+0.4}_{-0.4}$  & $2.5^{+0.1}_{-0.1}$   \\
    &    &   &  &  & $7.0^{+0.3}_{-0.3}$   & $2.7^{+0.3}_{-0.3}$  & $3.50^{+0.13}_{-0.13}$   \\
    &    &   &  &  & $13.6^{+0.5}_{-0.7}$   & $1.7^{+0.5}_{-0.6}$  & $6.4^{+4.7}_{-5.2}$   \\
    15 & 30178-01-10-00 & 50862.65  & 1947  &  C & $0.084^{+0.007}_{-0.005}$ & $4.7^{+1.3}_{-1.0}$ & $2.0^{+2.1}_{-1.1}$  \\
    &    &   &  &  & $4.9^{+0.1}_{-0.1}$   & $2.7^{+0.6}_{-0.3}$  & $3.9^{+1.7}_{-1.0}$   \\
    &    &   &  &  & $7.1^{+0.2}_{-0.1}$   & $2.6^{+0.5}_{-0.4}$  & $4.4^{+1.5}_{-1.2}$   \\
    &    &   &  &  & $13.9^{+0.3}_{-0.2}$   & $2.7^{+0.6}_{-0.3}$  & $3.9^{+1.7}_{-1.0}$   \\
    16 & 30188-02-11-00 & 50862.71  & 3230  &  C & $0.085^{+0.003}_{-0.003}$ & $4.4^{+0.6}_{-0.5}$ & $3.1^{+1.2}_{-1.0}$  \\
    &    &   &  &  & $4.9^{+0.1}_{-0.1}$   & $2.7^{+0.4}_{-0.3}$  & $3.7^{+1.0}_{-0.8}$   \\
    &    &   &  &  & $7.1^{+0.1}_{-0.1}$   & $2.9^{+0.3}_{-0.3}$  & $3.5^{+0.7}_{-0.6}$   \\
    &    &   &  &  & $13.6^{+0.2}_{-0.2}$   & $2.7^{+0.4}_{-0.3}$  & $3.7^{+1.0}_{-0.8}$   \\
    17 & 30178-02-03-00(1) & 50862.77  & 6771 & B & $0.098^{+0.003}_{-0.003}$ & $3.3^{+0.3}_{-0.3}$ & $4.0^{+1.3}_{-1.3}$  \\
    &    &   &  &  & $5.38^{+0.07}_{-0.08}$   & $4.5^{+0.3}_{-0.4}$  & $2.1^{+0.3}_{-0.3}$   \\
    &    &   &  &  & $13.2^{+0.4}_{-0.4}$   & $4.5^{+0.3}_{-0.4}$  & $2.1^{+0.3}_{-0.3}$   \\
    18 & 30178-02-03-00(2) & 50862.91  & 2820 &  C & $0.080^{+0.004}_{-0.004}$ & $4.4^{+0.5}_{-0.5}$ & $3.0^{+1.1}_{-0.8}$  \\
    &    &   &  &  & $4.7^{+0.1}_{-0.1}$   & $2.4^{+0.3}_{-0.3}$  & $3.7^{+0.9}_{-0.8}$   \\
    &    &   &  &  & $7.02^{+0.06}_{-0.06}$   & $2.6^{+0.2}_{-0.2}$  & $6.1^{+1.2}_{-1.0}$   \\
    &    &   &  &  & $13.4^{+0.3}_{-0.3}$   & $2.2^{+0.1}_{-0.2}$  & $4.4^{+1.4}_{-1.1}$   \\
    19 & 30178-01-11-00 & 50863.69  & 1445  & D &  None &  None   &  None   \\
    20 & 30188-02-12-00 & 50863.71  & 2638  & D &  None &  None   &  None   \\
    21 & 30188-02-13-00 & 50864.18  & 1287  & D &  None &  None   &  None   \\
    22 & 30188-02-14-00 & 50864.32  & 1108  & D &  None &  None   &  None   \\
    23 & 30178-01-12-00 & 50864.62  & 1730  & D &  None &  None   &  None   \\
    24 & 30188-02-15-00 & 50865.04  & 1578  & D &  None &  None   &  None   \\
    25 & 30188-02-16-00 & 50865.31  &  946  & C & $0.050^{+0.007}_{-0.009}$   & $5.8^{+0.8}_{-1.0}$  & $1.3^{+0.6}_{-0.8}$   \\
    &    &  &   &  & $7.03^{+0.09}_{-0.09}$   & $2.8^{+0.2}_{-0.2}$  & $5.6^{+1.5}_{-1.2}$  \\
    26 & 30188-02-17-00 & 50866.64  & 1478  & D &  None &  None   &  None   \\ 
    \hline
     \end{tabular}
     }
    \label{table: table1}

\end{table*}

\begin{table*}
    \caption{Spectral fitting results. a: $\rm NH$ is the column density of hydrogen in units of $10^{22} \rm cm^{-2}$; b: $E_{\rm cut}$ is the electron coronal temperature in units of keV, we fix it at 300 when it is less constrained;  c:  $R_{\rm in}$  is the inner radius of disc in units of $R_{\rm g}$; e: $\rm F_{\rm powerlaw}$  is the unabsorbed ﬂux of the {\sc cutoffpl} model ( 2.5 $\sim$ 30 keV) in units of $10^{-9} \rm erg/cm^2/s$; f: $\rm F_{\rm reflection}$  is the unabsorbed ﬂux of the {\sc relxill} model ( 2.5 $\sim$ 30 keV) in units of $10^{-9} \rm erg/cm^2/s$.}
    
    \scalebox{0.90}{
        \begin{tabular}{llllllllllll}
        \hline
        \multicolumn{1}{|l|}{\#} &
        \multicolumn{1}{l|}{ObsID} &
        \multicolumn{1}{l|}{MJD} &
        \multicolumn{1}{l|}{QRM} &
        \multicolumn{1}{l|}{$\chi^2_{\rm red}$} &
        \multicolumn{1}{l|}{$\rm NH^{\rm a}$} &   
        \multicolumn{1}{l|}{$\Gamma$} &
        \multicolumn{1}{l|}{$ E^{\rm b}_{\rm cut}$ } &
        \multicolumn{1}{l|}{$ R^{c}_{\rm in} $} &
        \multicolumn{1}{l|}{log$\xi$} &
        \multicolumn{1}{l|}{$\rm F^{\rm d}_{\rm powerlaw}$ } &
        \multicolumn{1}{l|}{$\rm F^{\rm e}_{\rm reflection}$ } \\ \hline
        1  & 30188--02--04--00  & 50856.97 & No & 1.61 & $15.0^{+0.2}_{-0.2}$   & $2.35^{+0.03}_{-0.01}$ & $300^{\rm fixed}$ & $5.2^{+30.7}_{-1.4}$ &  $> 4.4$ & $12.5^{+0.2}_{-0.2}$ & $3.0^{+0.2}_{-0.2}$  \\
        2  & 30178--01--05--00  & 50857.12 & Yes & 1.02 & $15.7^{+0.2}_{-0.2}$ & $2.40^{+0.02}_{-0.01}$ & $300^{\rm fixed}$ &  $4.8^{+6.6}_{-0.9}$ & $> 4.4 $ & $13.0^{+0.2}_{-0.2}$ & $3.4^{+0.2}_{-0.2}$\\
        3  & 30188--02--05--00  & 50857.71 & Yes & 0.49  & $16.7^{+0.2}_{-0.2}$ & $2.59^{+0.01}_{-0.01}$ & $300^{\rm fixed}$ & $4.2^{+1.9}_{-0.9}$ & $> 4.2$ & $14.3 ^{+0.1}_{-0.1}$ & $2.1 ^{+0.1}_{-0.1}$\\
        4  & 30178--02--02--00  & 50857.79 & Yes & 0.68 & $16.0^{+0.1}_{-0.1}$ & $2.55^{+0.01}_{-0.01}$  & $300^{\rm fixed}$ & $4.4 ^{+1.7}_{-0.8}$ & $> 4.5$ & $12.9^{+0.1}_{-0.1}$ & $2.4^{+0.1}_{-0.1}$\\
        5  & 30188--02--06--00  & 50858.05 & Yes & 1.02 & $15.6^{+0.2}_{-0.2}$ & $2.48^{+0.02}_{-0.01}$ & $300^{\rm fixed}$ & $3.6^{+2.0}_{-1.4}$ & $>4.5$ & $13.5^{+0.1}_{-0.1}$ & $3.0^{+0.1}_{-0.2}$\\
        6  & 30178--01--06--00  & 50858.70 & Yes & 1.01 & $15.6^{+0.2}_{-0.2}$ & $2.49^{+0.02}_{-0.01}$ & $300^{\rm fixed}$ &  $3.4^{+2.7}_{-2.3}$ & $>4.4$ & $13.6^{+0.2}_{-0.2}$ & $2.8^{+0.2}_{-0.2}$\\
        7  & 30188--02--07--00  & 50828.72 & Yes & 0.92  &  $15.6^{+0.1}_{-0.2}$ & $2.51^{+0.02}_{-0.01}$ & $300^{\rm fixed}$ & $3.1^{+2.2}_{-1.2}$ & $>4.3$ & $13.9^{+0.2}_{-0.1}$ & $2.5^{+0.1}_{-0.2}$\\
        8  & 30178--02--02--01  & 50858.77 & Yes & 0.58 & $15.7^{+0.1}_{-0.1}$ & $2.50^{+0.01}_{-0.01}$ & $300^{\rm fixed}$ & $4.1^{+1.8}_{-0.7}$ & $>4.5$   & $13.9^{+0.1}_{-0.1}$ & $2.8^{+0.1}_{-0.1}$\\
        9  & 30178--01--07--00  & 50859.84 & Yes & 0.74 & $16.7^{+0.2}_{-0.2}$ & $2.63^{+0.03}_{-0.01}$ & $300^{\rm fixed}$ & $3.3^{+1.4}_{-2.2}$ & $4.5^{+0.2}_{-0.6}$ & $15.7^{+0.1}_{-0.1}$ & $1.64^{+0.1}_{-0.1}$ \\
        10 & 30188--02--08--00  & 50860.12 & Yes & 0.68  & $17.0^{+0.2}_{-0.2}$ & $2.65^{+0.02}_{-0.01}$ & $300^{\rm fixed}$ &$2.5^{+1.2}_{-1.1}$ & $>4.2$ & $15.7^{+0.1}_{-0.1}$  & $2.2^{+0.1}_{-0.1}$\\
        11 & 30188--02--09--00  & 50860.56 & Yes & 0.80  & $16.0^{+0.2}_{-0.2}$ &  $2.57^{+0.02}_{-0.01}$ & $300^{\rm fixed}$ & $4.8^{+9.4}_{-1.5}$ & $>4.3$ & $15.5^{+0.2}_{-0.1}$ & $1.8^{+0.1}_{-0.2}$\\
        12 & 30178--01--08--00  & 50850.72 & Yes & 0.80 & $15.3^{+0.2}_{-0.2}$ & $2.51^{+0.02}_{-0.01}$ & $300^{\rm fixed}$ & $3.2^{+3.1}_{-1.5}$ & $>4,3$ & $14.8^{+0.2}_{-0.2}$ & $2.58^{+0.2}_{-0.2}$\\
        13 & 30178--01--09--00  & 50861.69 & Yes & 0.84 & $15.6^{+0.3}_{-0.3}$ & $2.49^{+0.04}_{-0.04}$ & $189.8^{+160.6}_{-62.3}$ & $5.2^{+5.3}_{-4.0}$ &  $4.5^{+0.2}_{-0.5}$ & $15.8^{+0.2}_{-0.1}$ & $1.78^{+0.1}_{-0.2}$\\
        14 & 30188--02--10--00  & 50861.72 & Yes &  0.56 & $15.7^{+0.2}_{-0.2}$ & $2.53^{+0.03}_{-0.01}$ & $300^{\rm fixed}$ & $3.7^{+4.3}_{-1.3}$ & $>3,6$ & $15.2^{+0.2}_{-0.2}$ & $2.3^{+0.2}_{-0.2}$\\
        15 & 30178--01--10--00  & 50862.65 & Yes &  0.64 & $15.4^{+0.3}_{-0.2}$ & $2.51^{+0.04}_{-0.03}$ & $174.9^{+103.5}_{-57.9}$ & $4.62^{+7.3}_{-1.1}$ & $>4,3$ & $14.9^{+0.2}_{-0.1}$ & $2.7^{+0.1}_{-0.2}$\\
        16 & 30188--02--11--00  & 50862.71 & Yes & 0.95 & $15.7^{+0.1}_{-0.1}$ & $2.52^{+0.01}_{-0.01}$ & $300^{\rm fixed}$ & $4.0^{+2.4}_{-2.7}$ & $4.6^{+0.1}_{-0.2}$ & $15.3^{+0.1}_{-0.1}$ & $2.3^{+0.1}_{-0.1}$\\
        17 & 30178--02--03--00(1)  & 50862.77 & Yes  & 0.95 & $15.5^{+0.7}_{-0.1}$ & $2.50^{+0.06}_{-0.01}$ & $300^{\rm fixed}$ & $3.5^{+2.3}_{-2.1}$ & $>2.5$ & $15.3^{+0.1}_{-0.1}$ & $2.0^{+0.1}_{-0.1}$\\ 
        18 & 30178--02--03--00(2)  & 50862.91   & Yes &  0.88     &  $15.9^{+0.4}_{-0.2}$ & $2.53^{+0.06}_{-0.02}$ & $145.3^{+155.7}_{-22.7}$ & $4.0^{+1.8}_{-1.8}$ & $>4.1$ & $15.4^{+0.1}_{-0.1}$ & $2.0^{+0.1}_{-0.1}$\\ 
        19 & 30178--01--11--00  & 50863.69   & No    & 1.24   & $18.1^{+0.2}_{-0.2}$ & $2.76^{+0.02}_{-0.02}$ & $98.1^{+10.3}_{-12.1}$ &$2.9^{+0.5}_{-0.9}$ & $>4.6$ & $19.5^{+0.2}_{-0.1}$ & $3.3^{+0.1}_{-0.2}$\\ 
        20 & 30188--02--12--00  & 50863.71   & No   & 1.40 & $18.1^{+0.1}_{-0.2}$ & $2.75^{+0.01}_{-0.03}$ & $99.4^{+3.6}_{ -21.0}$ & $2.7^{+0.8}_{-0.9}$ & $>4.5$ & $19.8^{+0.1}_{-0.1}$ & $3.2^{+0.1}_{-0.1}$\\
        21 & 30188--02--13--00  & 50864.18   & No    & 0.92 &  $16.9^{+0.3}_{-0.2}$ & $2.55^{+0.04}_{-0.02}$ & $73.9^{+16.5}_{-9.3}$ & $2.1^{+1.1}_{-0.5}$ & $4.6^{+0.1}_{-0.4}$ & $21.0^{+0.2}_{-0.2}$ & $3.3^{+0.2}_{-0.2}$\\ 
        22 & 30188--02--14--00  & 50864.32   & No   & 1.00 & $17.0^{+0.3}_{-0.3}$ & $2.57^{+0.04}_{-0.03}$ & $57.1^{+10.2}_{-7.9}$  & $2.3^{+1.6}_{-0.8}$ & $4.6^{+0.1}_{-0.4}$ & $20.1^{+0.2}_{-0.2}$ & $3.2^{+0.2}_{-0.2}$\\ 
        23 & 30178--01--12--00  & 50864.62   & No    & 0.99 & $18.2^{+0.8}_{-0.2}$  & $2.75^{+0.10}_{-0.02}$ & $74.8^{+14.0}_{-9.5}$ & $2.2^{+0.6}_{-1.2}$ & $>2.1$ & $19.9^{+0.2}_{-0.1}$ & $3.3^{+0.1}_{-0.1}$\\
        24 & 30188--02--15--00  & 50865.04   & No   & 0.97 & $17.0^{+0.3}_{-0.3}$ & $2.54^{+0.06}_{-0.04}$ & $60.0^{+17.4}_{-8.9}$ & $2.1^{+1.0}_{-0.5}$ & $4.6^{+0.1}_{-0.4}$ & $20.7^{+0.2}_{-0.2}$ & $3.6^{+0.2}_{-0.2}$\\
        25 & 30188--02--16--00  & 50865.31   & Yes    & 0.79 & $16.3^{+0.2}_{-0.1}$ & $2.64^{+0.02}_{-0.01}$ & $193.3^{+44.3}_{-38.8}$ & $2.1^{+1.1}_{-0.6}$ & $>4.4$ & $17.6^{+0.2}_{-0.2}$ & $2.1^{+0.2}_{-0.2}$\\ 
        26 & 30188--02--17--00  & 50866.64   &  No   & 1.15 & $17.8^{+0.4}_{-0.2}$ &  $2.64^{+0.05}_{-0.01}$ & $55.0^{+11.6}_{-6.7}$ & $2.2^{+6.5}_{-0.6}$ & $>4.4$ & $20.4^{+0.2}_{-0.2}$ & $3.7^{+0.2}_{-0.2}$\\ 
        \hline
         \end{tabular}
         }
         \label{table:table2}

\end{table*}

%%%%%%%%%%%%%%%%%%%%%%%%%%%%%%%%%%%%%%%%%%%%%%%%%%

% Don't change these lines
\bsp	% typesetting comment
\label{lastpage}
\end{document}